\documentclass{IEEEtran}
\usepackage{subfigure}
\usepackage[cmex10]{amsmath}
\usepackage{amssymb}
\usepackage{epsfig}
\usepackage{cite}
\usepackage{graphicx}
\usepackage{color}
\usepackage{bm}
\usepackage{booktabs}
\usepackage{gensymb}
\usepackage{mathrsfs}
\usepackage{xfrac}
\usepackage{algorithm}
\usepackage{algorithmic}
\usepackage{multirow}
\usepackage{float}
\newfloat{figtab}{htb}{fgtb}
\makeatletter
 \newcommand\figcaption{\def\@captype{figure}\caption}
  \newcommand\tabcaption{\def\@captype{table}\caption}

\newlength{\figwidth}
\newcommand{\tabincell}[2]{\begin{tabular}{@{}#1@{}}#2\end{tabular}}

\newtheorem{remark}{\it Remark}
\newtheorem{proposition}{\it Proposition}


\begin{document}
\title{Dynamic MIMO Architecture Design for \\ Near-Field Communications}
\author{Zheng Zhang,~\IEEEmembership{Graduate Student Member,~IEEE},  Yuanwei Liu,~\IEEEmembership{Fellow,~IEEE}, \\ Zhaolin Wang,~\IEEEmembership{Graduate Student Member,~IEEE}, Jian Chen,~\IEEEmembership{Member,~IEEE}, and Tony Q.S. Quek,~\IEEEmembership{Fellow,~IEEE} \vspace{-5mm}
\thanks{Zheng Zhang and Jian Chen are with the School of Telecommunications Engineering, Xidian University, Xi'an 710071, China (e-mail: zzhang\_688@stu.xidian.edu.cn; jianchen@mail.xidian.edu.cn).}
\thanks{Yuanwei Liu and Zhaolin Wang are with the School of Electronic Engineering and Computer Science, Queen Mary University of London, London E1 4NS, U.K. (e-mail: yuanwei.liu@qmul.ac.uk; zhaolin.wang@qmul.ac.uk;).}
\thanks{Tony Q. S. Quek is with the Singapore University of Technology and Design, Singapore 487372, and also with the Yonsei Frontier Lab, Yonsei University, South Korea (e-mail: tonyquek@sutd.edu.sg)}}

\maketitle

\begin{abstract}
  A novel dynamic hybrid beamforming architecture is proposed to achieve the spatial multiplexing-power consumption tradeoff for near-field multiple-input multiple-output (MIMO) networks, where each radio frequency (RF) chain is connected to each antenna using a couple of independent phase shifters to reduce the number of required RF chains. Based on this architecture, an optimization problem is formulated that maximizes the sum of achievable rates while minimizing the hardware power consumption. Both continuous and discrete phase shifters are considered. \textit{1) For continuous phase shifters}, a weighted minimum mean-square error-based two-stage (WMMSE-TS) algorithm is proposed, where the same performance as the optimal fully-digital beamformer can be achieved by the proposed hybrid beamformer even if number of RF chains equals the number of data streams. \textit{2) For discrete phase shifters}, a penalty-based layered iterative (PLI) algorithm is proposed. The closed-form analog and baseband digital beamformers are derived in each iteration. Simulation results demonstrate that: 1) the proposed dynamic beamforming architecture outperforms the conventional fixed hybrid beamforming architecture in terms of spatial multiplexing-power consumption tradeoff, and 2) the proposed algorithms achieve better performance than the other baseline schemes.
\end{abstract}
\IEEEpeerreviewmaketitle\vspace{-5mm}

\section{Introduction}\label{Section_1}
Driven by the ever-increased demand for ultra-high data rates and seamless wireless connectivity, multiple-input multiple-output (MIMO) has been extensively studied over the past decades \cite{Z.Zhang_6G,E.G.Larsson_MIMO}. As a key enabler for the fifth-generation (5G) wireless networks, MIMO encourages the deployment of multiple independently-controlled antenna arrays on transceivers and provides an enriched spatial degree-of-freedom (DoF) toward network capacity enhancement. To elaborate, by sending multiple data streams from the individual arrays of the base station (BS), MIMO imports more independent spatial dimensions for the propagation of information-bearing signals. On this basis, the simple phase coherence-based signal processing technique can be leveraged to adjust the electromagnetic (EM) wave propagation characteristics, where the wavefronts are coherently superimposed on the position of interest while destructively canceled on the other positions. Thus, MIMO enables the network throughput to rise linearly with the number of transmit/receive antennas without requiring extra transmit power or other time/frequency resources \cite{MIMO_linear} (referred as to \textit{spatial multiplexing gain}), which has been widely incorporated into a variety of wireless communication standards, such as fourth-generation (4G) long-term evolution (LTE) \cite{4G_LTE}, 5G new radio (NR) \cite{5G_NR}, and WiMAX \cite{WiMAX}.

To cope with more stringent communication requirements in emerging application scenarios of the beyond-5G era, e.g., extended reality (XR), remote healthcare, and auto-driving, MIMO networks are anticipated to scale up an extremely large-scale size to further enhance the spatial multiplexing gain \cite{M.Cui_NF_mag}. However, the significant enlargement of the number of antennas is generally accompanied by an increased antenna aperture, which extends the Fresnel (near-field) region of MIMO channels from a trivial distance to dozens or even hundreds of meters \cite{C.You_NFC_tutorial} . In the conventional Fraunhofer (far-field) region, the EM propagation can be rationally approximated as the planar wave, with the phase responses linearly related to the angles. Nevertheless, the signal propagation is fundamentally altered in near fields, where the accurate spherical-wave model is required to characterize the nonlinear phase variation concerning the corresponding angle and distance \cite{Y.Liu_NFC_tutorial}. Note that as the additional distance-domain information is introduced to the near-field array responses, it entails new opportunities for wireless communications. To elaborate, the unique near-field spherical-wave channels render it possible to distinguish the different locations in the same direction, which favors the focusing of the signal energy on free-space points (referred as to \textit{beamfocusing}), thus mitigating inter-user interference of multi-user communications \cite{H.Zhang_NFC_mag}. On the other hand, the wave curvatures across individual arrays are not identical in near fields, which inevitably leads to an increased rank of the line-of-sight (LoS) channels and provides a higher spatial multiplexing gain for network capacity enhancement \cite{NF_EDoF_mag}. Therefore, near-field communications have gained tremendous attention recently \cite{Y.Liu_NFC_mag}.
\vspace{-3mm}
\subsection{Prior Works}\vspace{-1mm}
The research on MIMO technology commenced with point-to-point systems. Specifically, the authors of \cite{R.Heath_MIMO} proposed a dynamic transmission scheme, which can switch transmission modes based on the instantaneous channel feedback, so as to exploit the spatial multiplexing gain and diversity gain fully. Followed by this, the work of \cite{MIMO_TIT} further demonstrated the possibility of concurrently obtaining spatial multiplexing gain and diversity gain in point-to-point MIMO networks and revealed the optimal multiplexing-diversity tradeoff. To accelerate the commercialization of MIMO technology, the research emphasis has turned to more practical multi-user MIMO (MU-MIMO) networks in recent years. In the work of \cite{E.Bjrnson_MIMO}, the authors investigated the uplink and downlink transmission of a MU-MIMO network. In particular, a novel refined power consumption model was proposed to characterize the realistic power scaling, which validated that MIMO was an energy-efficient technique for enhancing connectivity. As a further advance, the authors of \cite{S.Huberman_MIMO} developed a full-duplex-based beamforming transceiver structure, which can provide higher spectral efficiency by eliminating inter-user interference. To support high-frequency band (e.g., millimeter-wave band) MIMO communications, the phase-shift-based hybrid beamforming architecture is proposed for multi-stream transmission with less number of radio frequency (RF) chains \cite{Hybrid_O.E.Ayach}.

More recently, near-field communications have garnered widespread attention due to the dramatically increased antenna arrays scale. The authors of \cite{M.Cui_NFC_channel} pioneered the modeling of near-field channels for extremely large-scale array scenarios, where the polar-domain representation of spherical-wave channels was derived. Considering the cross-field scenarios, the authors of \cite{Z.Hu_hybrid_field} proposed an orthogonal matching pursuit (OMP) approach to estimate the far-field and near-field channels. For the multipath scenarios, the authors of \cite{Z.Yuan_NFC} proposed a spatial non-stationary-based channel modeling framework for the uniform circular array to characterize the physical propagation in near fields. Following this, a unified analysis framework oriented towards near-field channels with spatial non-stationary components was proposed in the work of \cite{NF_analysis}, where the effects of discrete array aperture and polarization mismatch on the received power were unveiled. Furthermore, the authors of \cite{LDMA} analyzed the asymptotic orthogonality in the distance domain of near-field channels, where an innovative concept of location division multiple access (LDMA) was proposed to exploit the distance-domain spatial DoF.

To further fulfill the potential of near-field communications, lots of endeavors have been devoted to the beamforming (also referred as to beamfocusing) design of near-field MIMO networks. From the interference-mitigation perspective, the authors of \cite{H.Zhang_NF} proposed to exploit the near-field beamfocusing effect to provide multiple orthogonal links in the same direction. To counter the uncertain near-field signal propagation, the work of \cite{AI_NF_MIMO} designed a tailored artificial intelligence (AI)-native transceiver framework to achieve robust beamfocusing. As a further advance, the work of \cite{X.Li_NF_MIMO} considered a modular array antenna structure, where the joint beamfocusing pattern and the user grouping optimization scheme were designed to provide higher spatial resolution. The authors of \cite{Z.Wang_NF_MIMO} proposed a novel sub-connected true-time delayer (TTD) hybrid beamforming structure for near-field wideband networks, which can inhibit the beam-split effect in near fields and improve the network energy efficiency. To fully utilize the spatial DoF provided by near-field channels, the work of \cite{Z.Wu_NF_MIMO} proposed a distance-aware hybrid beamforming architecture, which can effectively improve the MIMO network capacity.

\subsection{Motivations and Contributions}
Although there have been a few researches on near-field MIMO communications \cite{H.Zhang_NF,AI_NF_MIMO,X.Li_NF_MIMO,Z.Wang_NF_MIMO,Z.Wu_NF_MIMO}, the tailored beamforming strategy from the perspective of DoF is still in the initial stage. Actually, the most intuitive alteration brought by the near-field signal propagation to MIMO networks is the significantly enhanced channel rank, which leads to an improvement of the spatial multiplexing gain. However, the existing work oriented towards multiplexing gain utilization \cite{Z.Wu_NF_MIMO} only considered the sub-connected hybrid structure and a single-user scenario, which leads to an inferior performance to the fully-digital beamforming and cannot directly expand to the MU-MIMO networks. To fill this research gap, this paper considers a near-field MU-MIMO network, based on the following two important observations.
\begin{itemize}
  \item The high spatial multiplexing gain in the near fields also imposes huge hardware power consumption (HPC) and heavy hardware implementation complexity at the BS. Take an example of a conventional fully-connected hybrid beamforming structure, it requires twice the number of RF chains as the number of data streams to achieve the same performance as fully-digital beamforming, where the number of active devices (e.g., amplifiers, mixers, and transmitters) in RF chains and unit-modulus phase shifters becomes extremely large. Hence, it is crucial to design a new hybrid beamforming architecture that can balance the near-field spatial multiplexing gain and the HPC at the BS.
  \item Considering MU-MIMO scenarios, since the DoF of the near-field MIMO channel is distance-dependent, multiple users located at different distances from the BS usually enjoy different spatial multiplexing gains. It indicates that the conventional fixed data-stream allocation scheme in far fields will be incompatible with the near-field MIMO transmission. Therefore, an adaptive data-stream allocation strategy is required to flexibly allocate data streams emitted to different users for coinciding with their respective channel DoF.
\end{itemize}

\begin{figure*}[t]
  \centering
  \includegraphics[scale = 0.45]{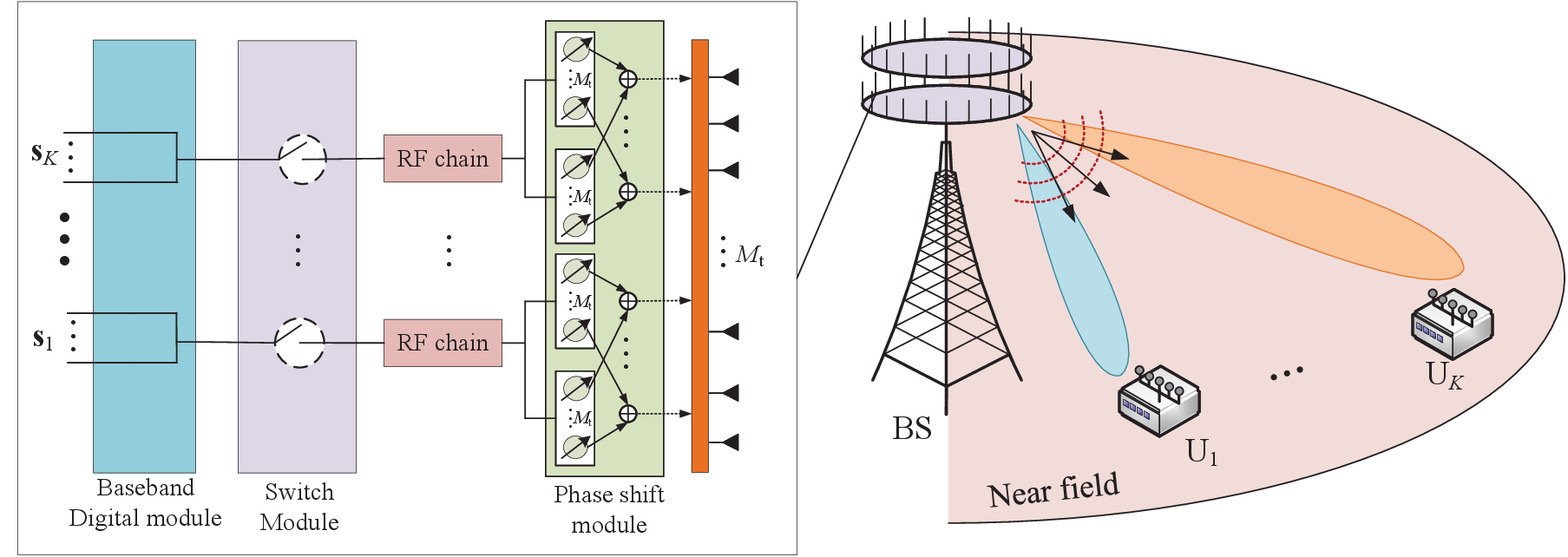}
  \caption{The near-field MIMO network with dynamic hybrid beamforming architecture.}\vspace{-5mm}
  \label{Fig.1}
\end{figure*}

Motivated by the above, this paper proposes a near-field MU-MIMO communication framework, where a BS is equipped with extremely large-scale array antennas to communicate with multiple multi-antenna users located in the near-field region, with particularly focusing on the tradeoff between communication sum rates and HPC. The main contributions of this paper are summarized as follows.
\begin{itemize}
  \item We propose a dynamic hybrid beamforming architecture, where a switch module is integrated between the baseband digital and analog phase-shift modules to adjust the number of activated RF chains and corresponding phase shifters. To achieve the same performance as fully-digital beamforming while keeping the number of RF chains at a minimum, each RF chain is connected to an arbitrary antenna via two independent phase shifters. Based on this architecture, we formulate a dynamic hybrid beamforming optimization problem that characterizes the rate-HPC tradeoff, where both ideal continuous phase shifters and practical discrete phase shifters are considered.
  \item For continuous phase shifters, we propose a weighted minimum mean-square error-based two-stage (WMMSE-TS) algorithm. To elaborate, we first employ the WMMSE approximation to optimize the fully-digital beamformer in an iterative manner. To facilitate the optimization of integer variables in the data-selection matrix, the contributions of different fully-digital beamforming vectors to the objective function are derived in the closed-form expressions. Then, the QR decomposition is carried out the extract the baseband digital and analog beamformers from the optimized fully-digital beamformer, which achieves the same performance as the fully-digital beamformer using as many RF chains as the number of data streams.
  \item For the practical discrete phase shift, we propose a penalty-based layered iterative (PLI) optimization framework, where the hybrid beamformer is integrated into the penalty term of the objective function. To characterize the impact of the data selection on the hybrid beamformer optimization, a summation-form-based penalty function is constructed, where the data-selection matrix serves as the weight of the hybrid beamformer. Based on this transformation, the block coordinate descent (BCD) method is employed to iteratively optimize the hybrid beamformers, where the analog phase shifters and baseband digital beamformer are updated in the closed-form expressions.
  \item Numerical results demonstrate the effectiveness of the proposed algorithms. It is also verified that: 1) the proposed dynamic hybrid beamforming architecture achieves better performance than the other baseline schemes; 2) the proposed algorithm can enable a flexible tradeoff between the communication rates and HPC; 3) 4-bit discrete phase shifters are sufficient to achieve a comparable performance of the ideal continuous phase shifters.
\end{itemize}

\vspace{-4mm}
\subsection{Organization and Notations}\vspace{-1mm}
The organization of this paper is as follows. Section \label{Section_2} introduces the system model and problem formulation. In Section \ref{Section_3}, the WMMSE-TS algorithm is proposed to optimize the hybrid beamforming under the continuous phase shift assumption. Section \ref{Section_4} conceives a PTL algorithm for the joint optimization of analog and baseband digital beamformers with consideration of the discrete phase shifts. Section \ref{Section_5} presents the numerical results. Finally, the conclusion is obtained in Section \ref{Section_6}.

\textit{Notations:} The scalar, vector, and matrix are denoted by the lower-case letter, boldface lower-case letter, and boldface capital, respectively. The complex matrix with $N$ rows and $M$ columns is denoted by $\mathbf{X}\in\mathbb{C}^{N\times M}$. The transpose and Hermitian conjugate operation of matrix $\mathbf{X}$ is denoted by $\mathbf{X}^{T}$ and $\mathbf{X}^{H}$. The $i$-th row and $j$-th column element of the matrix $\mathbf{X}$ is denoted by $\mathbf{X}^{[i,j]}$. The Euclidean distance between the scalar $x$ and $y$ is denoted by $|x-y|$, and the Euclidean norm of the vector $\mathbf{x}$ is denoted by $\|\mathbf{x}\|$. $\mathbf{x}\sim \mathcal{CN}(0,\mathbf{X})$ denotes a circularly symmetric complex Gaussian (CSCG) distributed vector with zero mean and covariance matrix $\mathbf{X}$. The trace and rank of the $\mathbf{X}$ are denoted by $\text{Tr}(\mathbf{X})$ and $\text{rank}(\mathbf{X})$. The Frobenius norm of the $\mathbf{X}$ is denoted by $\|\mathbf{X}\|_{\text{F}}$. $\mathbf{X}^{-1}$ represents the inverse matrix of the $\mathbf{X}$. The real component of the matrix $\mathbf{X}$ is denoted by $\Re{(\mathbf{X})}$. The statistical expectation of the matrix $\mathbf{X}$ is denoted by $\mathbb{E}(\mathbf{X})$. The pseudo-inverse of the matrix $\mathbf{X}$ is denoted by $\mathbf{X}^{\dag}$.

\section{System Model and Problem Formulation}\label{Section_2}
As shown in Fig. \ref{Fig.1}, we consider a downlink near-field MIMO network, where an $M_{\text{t}}$-antenna BS transmit signals to $K$ $M_{\text{r}}$-antenna users (denoted by $\{\text{U}_{1},\cdots,\text{U}_{K}\}$). All the users are assumed to be located in the near-field region of the BS, i.e., the distance from the BS to $\text{U}_{k}$ is shorter than the Rayleigh distance $\frac{2(D_{\text{BS}}+D_{k})^{2}}{\lambda}$, where $D_{\text{BS}}$ and $D_{k}$ denote the antenna apertures of the BS and $\text{U}_{k}$, respectively. To adaptively exploit the DoFs of the near-field MIMO channels while conserving the power consumption of the RF chains, we propose a dynamic hybrid beamforming architecture in this paper. To elaborate, the BS is equipped with $M_{\text{t}}^{\text{RF}}$  ($M_{\text{t}}^{\text{RF}}\ll M_{\text{t}}$) RF chains to send multi-stream data to $K$ users by employing the hybrid beamforming structure. A switch module is integrated between the baseband digital module and analog module, which consists of $M_{\text{t}}^{\text{RF}}$ ON-OFF circuits to control the on/off status of the RF chains. Each ON-OFF circuit can switch between the on and off modes, which activates the corresponding RF chain in the ON mode while turning off the RF chain in the OFF mode. Based on the ON-FF status of RF chains, the baseband digital module sends multiple data streams with the same number of activated RF chains, where the data streams are going to be manipulated by the analog beamformer and emitted to the antenna for signal broadcasting. Thus, by adjusting the switch module, we can proactively control the number of emitted data streams and the power consumed by the RF chains and analog beamformer. Furthermore, each RF chain at the BS is connected to each antenna through two independent phase shifters. Such a structure is capable of realizing the optimal fully-digital performance using a minimum number of RF chains \cite{RF_chains}, which will be detailed in the following sections.  At the user end, the fully-digital combining architecture is assumed due to the relatively small number of antennas, 

\subsection{Near-Field Channel Model}

\begin{figure}[t]
  \centering
  \includegraphics[scale = 0.55]{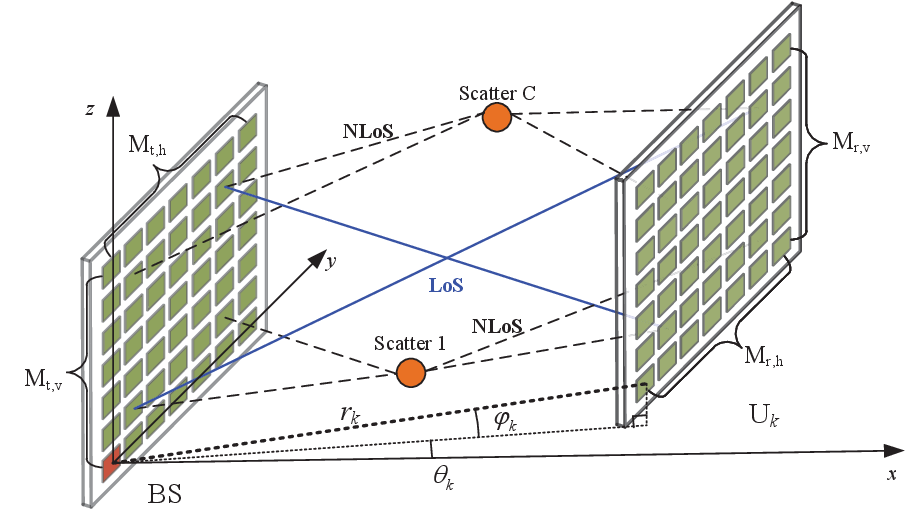}
  \caption{The multi-path UPA channel in near-field MIMO communications.}
  \label{Fig.2}
\end{figure}

Without loss of generality, multi-path MIMO channels are considered in this paper, where both the LoS and non-line-of-sight (NLoS) components are present due to the existence of multiple scatters in the near-field region. As shown in Fig. \ref{Fig.2}, the BS is assumed to be equipped with a uniform planar array (UPA) with $M_{\text{t}}=M_{\text{t},\text{v}}M_{\text{t},\text{h}}$ antennas, and $\text{U}_{k}$ is assumed to be equipped with a UPA with $M_{\text{r}}=M_{\text{r},\text{v}}M_{\text{r},\text{h}}$ antennas, where $M_{\text{t},\text{v}}$/$M_{\text{r},\text{v}}$ and $M_{\text{t},\text{h}}$/$M_{\text{r},\text{h}}$ denote the numbers of vertical antennas and horizontal antennas at the BS/$\text{U}_{k}$, respectively. A three-dimensional (3D) coordinate system is considered, where all the UPA planes are assumed to be located in the y-o-z plane. The origin of the coordinate system is put into the antenna at the first row and the first column of the UPA at the BS. Thus, the coordinate of the $v$-row and the $h$-column antenna of the BS is given by $\mathbf{s}_{v,h}=(0,\tilde{v}d,\tilde{h}d)$, where $d$ denotes the antenna spacing, $\tilde{v}=v-1$, and $\tilde{h}=h-1$. We further define $r_{k}$, $\theta_{k}$ and $\varphi_{k}$ as the distance, the azimuth angle, and the elevation angle of the $\text{U}_{k}$ with respect to the origin of the 3D coordinate system. The coordinate of the $m$-row and the $n$-column antenna of $\text{U}_{k}$ is given by $\mathbf{r}_{m,n}^{k}=(r_{k}\cos\theta_{k}\sin\varphi_{k},
r_{k}\sin\theta_{k}\sin\varphi_{k}+\tilde{m}d,r_{k}\cos\varphi_{k}+\tilde{n}d)$, where $\tilde{m}=m-1$ and $\tilde{n}=n-1$. Accordingly, the distance between the $v$-row and the $h$-column antenna of the BS and the $m$-row and the $n$-column antenna of $\text{U}_{k}$ is given by
\begin{align}\nonumber
&\|\mathbf{r}_{m,n}^{k} - \mathbf{s}_{v,h}\|=  \big[r_{k}^{2}+(\tilde{m}\!-\!\tilde{v})^{2}d^{2}+(\tilde{n}\!-\!\tilde{h})^{2}d^{2}+\\ \label{1}
&\qquad 2r_{k}\sin\theta_{k}\sin\varphi_{k}(\tilde{m}\!-\!\tilde{v})d+2r_{k}\cos\varphi_{k}(\tilde{n}\!-\!\tilde{h})d\big]^{\frac{1}{2}}.
\end{align}
Hence, the array response vector from the BS to the $m$-row and the $n$-column antenna of $\text{U}_{k}$ is given by
\begin{align}\nonumber
\mathbf{a}_{m,n}^{k}(r_{k},\theta_{k},\varphi_{k}) &= \big[e^{-\jmath\frac{2\pi}{\lambda}(\|\mathbf{r}_{m,n}^{k} - \mathbf{s}_{1,1}\|)},\cdots,\\ \label{2}
&\qquad\quad e^{-\jmath\frac{2\pi}{\lambda}(\|\mathbf{r}_{m,n}^{k} - \mathbf{s}_{M_{\text{t},\text{v}},M_{\text{t},\text{h}}}\|)}\big]^{T}.
\end{align}
Then, the LoS channel between the BS and $\text{U}_{k}$ can be expressed as
\begin{align}\nonumber
\mathbf{H}_{k}^{\text{LoS}} = \tilde{\beta}_{k}e^{-\jmath\frac{2\pi}{\lambda}r_{k}}\big[\mathbf{a}_{1,1}^{k}&(r_{k},\theta_{k},\varphi_{k}),\cdots,\\ \label{3}
&\qquad\quad\mathbf{a}_{M_{\text{r},\text{v}},M_{\text{r},\text{h}}}^{k}(r_{k},\theta_{k},\varphi_{k})\big]^{T},
\end{align}
where $\tilde{\beta}_{k}$ denotes the complex gain \cite{Y.Liu_NFC_tutorial}, and $f$ denotes the operation frequency. On the other hand, due to the existence of the scatters, the NLoS components between the BS and $\text{U}_{k}$ cannot be neglected, which can be modeled as the product of the array response vectors of the BS and $\text{U}_{k}$, i.e.,
\begin{align}\label{4}
\mathbf{H}_{k}^{\text{NLoS}} &= \sum_{l=1}^{L}\tilde{\beta}_{l}\mathbf{a}_{l}^{k}(r_{l},\theta_{l},\varphi_{l})(\mathbf{a}_{l}^{\text{BS}}
(r_{l},\theta_{l},\varphi_{l}))^{T},
\end{align}
where $\tilde{\beta}_{l}$ denotes the corresponding complex channel gain with a random phase that is independent
and identically distributed and uniformly distributed in $[-\pi,\pi)$ and $L$ denotes the number of scatters in the considered network. $r_{l}$, $\theta_{l}$, and $\varphi_{l}$ denote the distance, the azimuth angle, and the elevation angle of the $l$-th scatter with respect to the coordinate $(0,0,0)$. The receive array response vector $\mathbf{a}_{l}^{k}(r_{l},\theta_{l},\varphi_{l})$ and the transmit array response vector $\mathbf{a}_{l}^{\text{BS}}(r_{l},\theta_{l},\varphi_{l})$ are given by
\begin{align}\nonumber
 \mathbf{a}_{l}^{k}(r_{l},\theta_{l},\varphi_{l}) = &\big[e^{-\jmath\frac{2\pi}{\lambda}(\|\mathbf{r}_{l} - \mathbf{r}_{1,1}^{k}\|)},\cdots,\\  \label{5}
 &\qquad e^{-\jmath\frac{2\pi}{\lambda}(\|\mathbf{r}_{l} - \mathbf{r}_{M_{\text{r},\text{v}},M_{\text{r},\text{h}}}^{k}\|)}\big]^{T},\\ \nonumber
 \mathbf{a}_{l}^{\text{BS}}(r_{l},\theta_{l},\varphi_{l}) = &\big[e^{-\jmath\frac{2\pi}{\lambda}(\|\mathbf{r}_{l} - \mathbf{s}_{1,1}\|)},\cdots,\\  \label{6}
 &\qquad e^{-\jmath\frac{2\pi}{\lambda}(\|\mathbf{r}_{l} - \mathbf{s}_{M_{\text{t},\text{v}},M_{\text{t},\text{h}}}\|)}\big]^{T},
\end{align}
where the coordinate of the $c$-th scatter is given by $\mathbf{r}_{l}=(r_{l}\cos\theta_{l}\sin\varphi_{l},
r_{l}\sin\theta_{l}\sin\varphi_{l},r_{l}\cos\varphi_{l})$. As a result, the near-field MIMO channel between the BS and $\text{U}_{k}$ is given by
\begin{align}\label{7}
\mathbf{H}_{k} = \mathbf{H}_{k}^{\text{LoS}}+\mathbf{H}_{k}^{\text{NLoS}}.
\end{align}
To characterize the fundamental performance boundary of the near-field MIMO network, we assume that the full CSI of users is available at the BS.

\subsection{Signal Model}
Based on the dynamic hybrid beamforming structure, the BS maximally transmits $M_{\text{t}}^{\text{RF}}$ data streams, and each user can receive a maximum of $M_{\text{r}}$-stream data. Thus, the transmitted data streams at the BS are given by
\begin{align}\label{8}
\mathbf{x}
=\sum_{k=1}^{K}\mathbf{P}\mathbf{W}_{k}\mathbf{T}_{k}\mathbf{s}_{k}.
\end{align}
where $\mathbf{P}\in\mathbb{C}^{M_{\text{t}}\times T_{\text{s}}}$ denotes the analog beamformer at the BS, $\mathbf{W}_{k}\in\mathbb{C}^{T_{\text{s}}\times M_{\text{r}}}$ denotes the transmit baseband digital beamformer of $\text{U}_{k}$, $\mathbf{T}_{k}\in\mathbb{C}^{M_{\text{r}}\times M_{\text{r}}}$ denotes the data-selection matrix for $\text{U}_{k}$, and $\mathbf{s}_{k}\in\mathbb{C}^{M_{\text{r}}\times 1}$ denotes the independent and identically distributed data symbols that $\text{U}_{k}$ can maximally receive. Here, $T_{\text{s}}$ denotes the number of the activated RF chains, which imposes the direct impact on the dimensions of $\mathbf{P}$ and
$\mathbf{W}_{k}$. To control the number of the transmitted data streams, $\mathbf{T}_{k}$ is defined as a diagonal matrix, which is given by
\begin{align}\label{9}
\mathbf{T}_{k} = \begin{bmatrix}
\mathbf{T}_{k}^{[1,1]}& \cdots & \mathbf{0}\\
\vdots &  \ddots & \vdots \\
\mathbf{0}&  \cdots & \mathbf{T}_{k}^{[M_{\text{r}},M_{\text{r}}]}
\end{bmatrix},
\end{align}
where $\mathbf{T}_{k}^{[m_{\text{r}},m_{\text{r}}]} \in \{0,1\}$ ($1\leq m_{\text{r}}\leq M_{\text{r}}$) is the integer variable. Also, the number of the transmitted data streams equals the number of the activated RF chains, which is denoted by $T_{\text{s}}=\sum_{k=1}^{K}\text{Tr}(\mathbf{T}_{k})$. For the analog beamformer, each entry of $\mathbf{P}$ satisfies the following constraint
\begin{align}\label{10}
\mathbf{P}^{[i,j]} = \mathbf{P}_{1}^{[i,j]}+\mathbf{P}_{2}^{[i,j]}, \  \forall i,j,
\end{align}
where $\mathbf{P}_{1}^{[i,j]}$ and $\mathbf{P}_{2}^{[i,j]}$ are the phase shifters with unit-modulus constraint. In this paper, two categories of phase shifters are considered:
\begin{itemize}
  \item \textbf{Continuous phase shifter:} In this case, the phase shifts in $\mathbf{P}_{1}$ and $\mathbf{P}_{2}$ are implemented by an infinite number of bits of phase shifters. Thus, $\mathbf{P}_{1}^{[i,j]}$ and $\mathbf{P}_{2}^{[i,j]}$ can be adjusted to the arbitrary value in the interval $[0,2\pi)$, i.e.,
  \begin{align} \label{11}
    \mathbf{P}_{i}\in\Phi_{1}\triangleq\{e^{\jmath\theta}|\theta\in[0,2\pi)\}, i\in\{1,2\}
\end{align}
  \item \textbf{Discrete phase shifter:} In this case, the phase shifts of $\mathbf{P}_{1}$ and $\mathbf{P}_{2}$ can only be taken over a limited number of discrete values, i.e.,
  \begin{align} \nonumber
    \mathbf{P}_{i}\in\Phi_{2}\triangleq&\left\{e^{\jmath\theta}|\theta\in\left\{0,\frac{2\pi}{2^{D}},\dots,\frac{2\pi(2^{D}-1)}{2^{D}}\right\}\right\}, \\ \label{12}
    &\qquad\qquad\qquad\qquad\qquad i\in\{1,2\},
\end{align}
where $D$ denotes the number of discrete bits.
\end{itemize}
Accordingly, the received signal at the $\text{U}_{k}$ is given by
\begin{align} \label{13}
\mathbf{y}_{k} = &\underbrace{\mathbf{H}_{k}\mathbf{P}\mathbf{W}_{k}\mathbf{T}_{k}\mathbf{s}_{k}}_{\text{desired signal}}+\underbrace{\sum_{j\neq k}^{K}\mathbf{H}_{k}\mathbf{P}\mathbf{W}_{j}\mathbf{T}_{j}\mathbf{s}_{j}}_{\text{interference}}+
\underbrace{\mathbf{n}_{k}}_{\text{noise}},
\end{align}
where $\mathbf{n}_{k}\sim\mathcal{CN}(0,\sigma^{2}\mathbf{I})$ denotes the additive white Gaussian noise (AWGN) at the $\text{U}_{k}$. The achievable rate of $\text{U}_{k}$ can be expressed as
\begin{align}\label{14}
R_{k} = \log_{2}\left|\mathbf{I}_{M_{\text{r}}}+\mathbf{Q}_{k}^{-1}\mathbf{H}_{k}\mathbf{P}\mathbf{W}_{k}\mathbf{T}_{k}\mathbf{T}_{k}^{H}
\mathbf{W}_{k}^{H}\mathbf{P}^{H}\mathbf{H}_{k}^{H}\right|,
\end{align}
where $\mathbf{Q}_{k}=\sum_{j\neq k}^{K}\mathbf{H}_{k}\mathbf{P}\mathbf{W}_{j}\mathbf{T}_{j}\mathbf{T}_{j}^{H}
\mathbf{W}_{j}^{H}\mathbf{P}^{H}\mathbf{H}_{k}^{H}+\sigma^{2}\mathbf{I}_{M_{\text{r}}}$ denotes interference plus noise covariance matrix at the $\text{U}_{k}$. Since introducing the switching module determines the number of employed RF chains and analog phase shifters, the HPC at the BS is given by
\begin{align}\label{15}
P_{\text{h}} = (P_{\text{RF}}+2M_{\text{t}}P_{\text{A}})T_{\text{s}},
\end{align}
where $P_{\text{RF}}$ and $P_{\text{A}}$ denote the power consumption of a single RF chain and single analog phase shifter.

\subsection{Problem Formulation}
In this paper, we focus on achieving the tradeoff between communication performance and the HPC. To this end, we formulate an optimization problem of maximizing the sum achievable rates of users while minimizing the HPC by jointly optimizing the data-selection matrix, baseband digital beamformer, and analog beamformer, subject to the total transmit power budget, the unit-modulus constraints in the analog beamformer, and the integer constraints of the data-selection matrix. The optimization problem is formulated as follows.
\begin{subequations}
\begin{align}
\label{16a} (\text{P}1)\  \max\limits_{\mathbf{P},\mathbf{W}_{k},\mathbf{T}_{k}}\   &\beta\left(\sum\nolimits_{k=1}^{K} R_{k}\right)- (1-\beta) P_{\text{h}}\\
\label{16b}\text{s.t.} \ \  & \sum_{k=1}^{K}\|\mathbf{P}\mathbf{W}_{k}\mathbf{T}_{k}\|^{2}_{F}\leq P_{\text{max}},\\
\label{16c}  &\mathbf{T}_{k}^{[m_{\text{r}},m_{\text{r}}]} \in\{0,1\}, \ \  \forall m_{\text{r}}, \forall k,\\
\label{16d}  &\mathbf{P}_{i}\in\Phi_{1}\  \text{or}\  \Phi_{2} , \quad \forall i,
\end{align}
\end{subequations}
where $\beta$ denotes the weight factor to adjust the importance of the communication rate and the hardware power consumption. \eqref{16b} limits the transmit power to be smaller than $P_{\text{max}}$; Constraints \eqref{16c} and \eqref{16d} account for the hardware constraints of the switch network and the analog beamformer, respectively. Note that the problem (P1) is a mixed integer nonlinear programming (MINLP) problem, which is non-deterministic polynomial (NP)-hard and is generally challenging to solve.

\begin{remark}
    Considering the same scattering environment with $L$ scatters, the DoF of the near-field channel between the BS and $\text{U}_{k}$ can be calculated according to
    $\text{DoF}_{k}=\min\left\{\frac{2(M_{\text{t},\text{v}}-1)(M_{\text{t},\text{h}}-1)
    (M_{\text{r},\text{v}}-1)(M_{\text{r},\text{h}}-1)d^{4}}{(\lambda r_{k})^{2}}\!+\!L,M_{\text{r}}\!+\!L,M_{\text{t}}\!+\!L\right\}$ \cite{DoFs}, which is inversely proportional to the distance $r_{k}$. However, the DoF of the far-field LoS channel is $1+L$. Thus, near-field MIMO channels lead to different DoFs for users positioned at different locations, and $\mathbf{T}_{k}$ should be introduced to adaptively adjust the number of the transmitted data streams for fully exploiting multiplexing gain, which, however, is constant in the far-field MIMO communications because the DoFs of far-field MIMO channels are distance-independent. As a result, the optimization problem (P1) is a unique problem in near-field MIMO communications.
\end{remark}

\section{Beamforming Optimization with Continuous Phase Shifters}\label{Section_3}
In this section, we concentrate on the hybrid beamforming design for the ideal continuous phase shift case, i.e., $\mathbf{P}_{i}\in\Phi_{1}$, $i\in\{1,2\}$. To deal with the non-convex problem, a WMMSE-TS algorithm is proposed. In the first stage, the WMMSE method is adopted to optimize the fully-digital beamformer and the data-selection matrices. Based on the optimized results, the closed-form hybrid beamformer is obtained in the second stage.

\subsection{Problem Formulation}

To facilitate the optimization, we introduce a Proposition to derive the optimal hybrid beamformers based on any given fully-digital beamformer.
\begin{proposition}\label{Proposition_1}
    With any optimized fully-digital beamformer $\mathbf{\bar{W}}=[\mathbf{\bar{W}}_{1},\cdots,\mathbf{\bar{W}}_{K}]$, the optimal analog and baseband digital beamformers are given by
    \begin{align}\label{17}
    \mathbf{\bar{W}} = \mathbf{P}\mathbf{W}_{\text{BB}},
    \end{align}
    where the baseband digital beamformer is given by $\mathbf{W}_{\text{BB}}=\mathbf{\Xi}\mathbf{\tilde{W}}_{2}^{H}\mathbf{\tilde{W}}_{1}$. Let $a_{[i,j]}$ denotes the amplitude of $\mathbf{P}^{[i,j]}$ and $\vartheta_{[i,j]}$ denotes the phase of $\mathbf{P}^{[i,j]}$, the unit-modulus phase shifts are given by
    \begin{align}\label{18}
\begin{cases}
        \mathbf{P}_{1}^{[i,j]}=e^{\jmath\left(\cos^{-1}\left(\frac{a_{[i,j]}}{2}\right)+\vartheta_{[i,j]}\right)},\\
        \mathbf{P}_{2}^{[i,j]}=e^{-\jmath\left(\cos^{-1}\left(\frac{a_{[i,j]}}{2}\right)-\vartheta_{[i,j]}\right)}.
        \end{cases}
\end{align}
\end{proposition}
\begin{IEEEproof}
See Appendix A.
\end{IEEEproof}

Let $\mathbf{\bar{W}}_{k}$ denotes the fully-digital beamformer for $\text{U}_{k}$, the optimization problem (P1) can be equivalently transformed into the following problem with respect to $\{\mathbf{\bar{W}}_{1},\cdots,\mathbf{\bar{W}}_{K}\}$
\begin{subequations}
\begin{align}
\label{19a} (\text{P}2)\  \max\limits_{\mathbf{\bar{W}}_{k},\mathbf{T}_{k}}\   &\beta\left(\sum\nolimits_{k=1}^{K} R_{k}\right)- (1-\beta) P_{\text{h}}\\
\label{19b}\text{s.t.} \ \  & \sum_{k=1}^{K}\|\mathbf{\bar{W}}_{k}\mathbf{T}_{k}\|^{2}_{F}\leq P_{\text{max}},\\
\label{19c}  &\mathbf{T}_{k}^{[m_{\text{r}},m_{\text{r}}]} \in\{0,1\}, \ \  \forall m_{\text{r}}, \forall k,
\end{align}
\end{subequations}
where $R_{k}= \log_{2}\big|\mathbf{I}_{M_{\text{r}}}+(\sum_{j\neq k}^{K}\mathbf{H}_{k}\mathbf{\bar{W}}_{j}\mathbf{T}_{j}\mathbf{T}_{j}^{H}
\mathbf{\bar{W}}_{j}^{H}\mathbf{H}_{k}^{H}+\sigma^{2}\mathbf{I}_{M_{\text{r}}})^{-1}\mathbf{H}_{k}\mathbf{\bar{W}}_{k}\mathbf{T}_{k}\mathbf{T}_{k}^{H}
\mathbf{\bar{W}}_{k}^{H}\mathbf{H}_{k}^{H}\big|$. However, the problem (P2) remains intractable to handle due to the non-convex achievable rate expression. To tackle this issue, we adopt the WMMSE approach to convert \eqref{19a} to an equivalent WMMSE expression \cite{PDD}. To elaborate, each receiver utilizes a combiner $\mathbf{Z}_{k}$ to estimate the received signals, and the mean-square error (MSE) matrix of $\text{U}_{k}$ can be defined by
\begin{align}\nonumber
&\mathbb{F}(\mathbf{Z}_{k},\mathbf{\bar{W}}_{k},\mathbf{T}_{k}) = \mathbf{Z}_{k}^{H}\mathbf{\dot{Q}}_{k}\mathbf{Z}_{k}+ \\ \label{20}
& (\mathbf{I}_{M_{\text{r}}}-\mathbf{Z}_{k}^{H}\mathbf{H}_{k}\mathbf{\bar{W}}_{k}\mathbf{T}_{k})
(\mathbf{I}_{M_{\text{r}}}-\mathbf{Z}_{k}^{H}\mathbf{H}_{k}\mathbf{\bar{W}}_{k}\mathbf{T}_{k})^{H},
\end{align}
where  $\mathbf{\dot{Q}}_{k}=\sum_{j\neq k}^{K}\mathbf{H}_{k}\mathbf{\bar{W}}_{j}\mathbf{T}_{j}\mathbf{T}_{j}^{H}
\mathbf{\bar{W}}_{j}^{H} \mathbf{H}_{k}^{H} +\sigma^{2}\mathbf{I}_{M_{\text{r}}}$. Then, the optimization problem (P2) is reformulated as
\begin{subequations}
\begin{align}
\nonumber (\text{P}2.1)\  \max\limits_{\mathbf{\bar{W}}_{k},\mathbf{\Gamma}_{k},
\atop
\mathbf{Z}_{k},\mathbf{T}_{k}}\  & \beta\Big(\sum_{k=1}^{K}
\Big[\log_{2}|\mathbf{\Gamma}_{k}|+\mu M_{\text{r}}\\ \nonumber
&-\mu\text{Tr}(\mathbf{\Gamma}_{k}\mathbb{F}(\mathbf{Z}_{k},\mathbf{\bar{W}}_{k},\mathbf{T}_{k}))\Big]\Big)\\ \label{21a}
&-(1-\beta) P_{\text{h}}\\
\label{21b}\text{s.t.} \quad & \eqref{19b},\eqref{19c},
\end{align}
\end{subequations}
where $\mathbf{\Gamma}_{k}\succeq\mathbf{0}$ is an auxiliary variable. Different from the conventional WMMSE approximation, we introduce a new scaling factor $\mu$, which is a constant coefficient that allows the MSE term to be of similar magnitude to the HPC. Notably, the optimization variables are separated in the constraints and coupled only in the convex/affine expressions, which inspires us to adopt the BCD optimization framework in the following.

\subsection{Proposed WMMSE-TS Algorithm}
In the first stage, we focus on the joint optimization of $\mathbf{\bar{W}}$ and the data-selection matrices.
\subsubsection{Subproblem with respect to $\{\mathbf{Z}_{k}\}$}
The problem (P2.1) is reduced to
\begin{subequations}
\begin{align}
\nonumber& (\text{P}2.2)  \min\limits_{\mathbf{Z}_{k}}\   \sum_{k=1}^{K}
\text{Tr}\left(\mathbf{\Gamma}_{k}\mathbf{Z}_{k}^{H}\left(\mathbf{H}_{k}\mathbf{\bar{W}}_{k}\mathbf{T}_{k}\mathbf{T}_{k}^{H}
\mathbf{\bar{W}}_{k}^{H}\mathbf{H}_{k}^{H}\right)\mathbf{Z}_{k}\right)\\ \label{22a}
&+\sum_{k=1}^{K}
\text{Tr}\left(\mathbf{\Gamma}_{k}\mathbf{Z}_{k}^{H}\mathbf{\dot{Q}}_{k}\mathbf{Z}_{k}\right)-
2\Re(\text{Tr}(\mathbf{\Gamma}_{k}\mathbf{Z}_{k}^{H}\mathbf{H}_{k}\mathbf{\bar{W}}_{k}\mathbf{T}_{k}))
\end{align}
\end{subequations}
The problem (P2.2) is a convex program, and the optimal $\mathbf{Z}_{k}$ can be derived via the first-order optimal condition, which is given by
\begin{align}\label{23}
\mathbf{Z}_{k} = \left(\mathbf{\dot{Q}}_{k}+\mathbf{H}_{k}\mathbf{\bar{W}}_{k}\mathbf{T}_{k}\mathbf{T}_{k}^{H}
\mathbf{\bar{W}}_{k}^{H}\mathbf{H}_{k}^{H}\right)^{-1}\mathbf{H}_{k}\mathbf{\bar{W}}_{k}\mathbf{T}_{k}.
\end{align}

\subsubsection{Subproblem with respect to $\{\mathbf{\Gamma}_{k}\}$} In this case, the problem (P2.1) is transformed into
\begin{subequations}
\begin{align}
\label{24a} \!\!\!(\text{P}2.3)\  \min\limits_{\mathbf{\Gamma}_{k}}\  & \sum_{k=1}^{K}
\big[\log_{2}|\mathbf{\Gamma}_{k}|\!-\!\mu\text{Tr}(\mathbf{\Gamma}_{k}\mathbb{F}(\mathbf{Z}_{k},\mathbf{\bar{W}}_{k},\mathbf{T}_{k}))\big]
\end{align}
\end{subequations}
Note that since $\mathbf{\Gamma}_{k}\succeq \mathbf{0}$ is a Hermitian matrix, the objective function is convex and the optimal $\mathbf{\Gamma}_{k}$ can be derived by
\begin{align}\label{25}
\mathbf{\Gamma}_{k} = \frac{\mathbb{F}(\mathbf{Z}_{k},\mathbf{\bar{W}}_{k},\mathbf{T}_{k})^{-1}}{\mu}.
\end{align}

\subsubsection{Subproblem with respect to  $\{\mathbf{\bar{W}}_{k},\mathbf{T}_{k}\}$} the optimization problem with respect to $\{\mathbf{\bar{W}}_{k},\mathbf{T}_{k}\}$ is formulated as
\begin{subequations}
\begin{align}\nonumber
 (\text{P}2.4)\  \min\limits_{\mathbf{\bar{W}}_{k},\mathbf{T}_{k}}\  & \sum_{k=1}^{K}
\Big[\beta\mu\text{Tr}(\mathbf{\Gamma}_{k}\mathbb{F}(\mathbf{Z}_{k},\mathbf{\bar{W}}_{k},\mathbf{T}_{k}))+\\ \label{26a}
&(1-\beta)(P_{\text{RF}}+2M_{\text{t}}P_{\text{A}})\text{Tr}(\mathbf{T}_{k})\Big]\\
\label{26b}\text{s.t.} \quad & \eqref{19b},\eqref{19c}.
\end{align}
\end{subequations}
Note that the problem (P2.4) is a MINLP problem and cannot be optimally solved unless using the exhaustive searching method. However, the exhaustive search method suffers an unacceptable computational complexity. To address this issue, we consider utilizing the Bisection method to effectively optimize $\{\mathbf{\bar{W}}_{k},\mathbf{T}_{k}\}$ with a low complexity. We first rewrite the problem (P2.4) as
\begin{subequations}
\begin{align}\nonumber
 \!\!\!\!(\text{P}2.5) \min\limits_{\mathbf{\bar{w}}_{j,k},\mathbf{T}_{k}} &\sum_{k=1}^{K}\sum_{j=1}^{M_{\text{r}}}\mathbf{T}_{k}^{[j,j]}\Big[\beta\mu\Big(-2\Re\Big(\mathbf{\bar{w}}_{j,k}^{H}\mathbf{f}_{j,k}\Big)+\\ \nonumber
&\mathbf{\bar{w}}_{j,k}^{H}\Big(
 \sum\nolimits_{i=1}^{K}\mathbf{H}_{i}^{H}\mathbf{Z}_{i}\mathbf{\Gamma}_{i} \mathbf{Z}_{i}^{H}\mathbf{H}_{i}\Big)
\mathbf{\bar{w}}_{j,k}\Big)\\ \label{27a}
&+(1-\beta)(P_{\text{RF}}+2M_{\text{t}}P_{\text{A}})\Big]\\
\label{27b}\text{s.t.} \quad & \sum_{k=1}^{K}\sum_{j=1}^{M_{\text{r}}}\mathbf{T}_{k}^{[j,j]}\|\mathbf{\bar{w}}_{j,k}\|^{2}\leq P_{\text{max}},,\\
\label{27c} & \eqref{19c},
\end{align}
\end{subequations}
where $\mathbf{F}_{k}=\mathbf{H}_{k}^{H}\mathbf{Z}_{k}\mathbf{\Gamma}_{k}=[\mathbf{f}_{1,k},\cdots, \mathbf{f}_{M_{\text{r}},k}]$, and $\mathbf{\bar{W}}_{k}=[\mathbf{\bar{w}}_{1,k},\cdots, \mathbf{\bar{w}}_{M_{\text{r}},k}]$. Given any $\mathbf{T}_{k}$, the optimal $\{\mathbf{\bar{w}}_{j,k}\}$ can be derived by
\begin{align}\label{28}
\mathbf{\bar{w}}_{j,k} =\left(\xi\mathbf{I}_{M_{\text{t}}}+\sum_{i=1}^{K}\mathbf{H}_{i}^{H}\mathbf{Z}_{i}\mathbf{\Gamma}_{i} \mathbf{Z}_{i}^{H}\mathbf{H}_{i}\right)^{-1}\mathbf{f}_{j,k},
\end{align}
where $\xi>0$  represents the Lagrangian multiplier of the Lagrangian function $\mathcal{L}(\mathbf{\bar{w}}_{j,k},\xi)=\sum_{k=1}^{K}\sum_{j=1}^{M_{\text{r}}}(\mathbf{\bar{w}}_{j,k}^{H}(
\sum_{i=1}^{K}\mathbf{H}_{i}^{H}\mathbf{Z}_{i}\mathbf{\Gamma}_{i} \mathbf{Z}_{i}^{H}\mathbf{H}_{i})\mathbf{\bar{w}}_{j,k})-2\Re(\mathbf{\bar{w}}_{j,k}^{H}\mathbf{f}_{j,k})+\xi(\sum_{k=1}^{K}\sum_{j=1}^{M_{\text{r}}}
\mathbf{\bar{w}}_{j,k}^{H}\mathbf{\bar{w}}_{j,k}^{H}-P_\text{max})$. Let $\mathbf{G}\mathbf{J}\mathbf{G}^{H}$ denote the eigenvalue decomposition (EVD) of the constant term $\sum_{i=1}^{K}\mathbf{H}_{i}^{H}\mathbf{Z}_{i}\mathbf{\Gamma}_{i} \mathbf{Z}_{i}^{H}\mathbf{H}_{i}$. Thus, the optimal $\xi$ can be obtained via the one-dimension search under the following equation satisfied.
\begin{align}\label{29}
\sum_{k=1}^{K}\sum_{j=1}^{M_{\text{r}}}\mathbf{T}_{k}^{[j,j]}\left(\sum_{i=1}^{M_{\text{t}}}\frac{x_{i,j,k}}{(\mathbf{J}^{[i,i]}+\xi)^{2}}\right)
=P_{\text{max}},
\end{align}
where $x_{i,j,k}=(\mathbf{G}^{H}\mathbf{f}_{j,k}\mathbf{f}_{j,k}^{H}\mathbf{G})^{[i,i]}$. Nevertheless, we cannot directly search $\xi$ due to the existence of the uncertain $\mathbf{T}_{k}$. In the following, we will show the value of $\mathbf{T}_{k}$ is dependent on the value of $\xi$.

Substituting \eqref{28} and $\sum_{i=1}^{K}\mathbf{H}_{i}^{H}\mathbf{Z}_{i}\mathbf{\Gamma}_{i} \mathbf{Z}_{i}^{H}\mathbf{H}_{i}\overset{\text{EVD}}{=}\mathbf{G}\mathbf{J}\mathbf{G}^{H}$ into \eqref{27a}, the objective function \eqref{27a} can be rewritten as
\begin{align}\nonumber
\eqref{27a} \!=\! &\sum_{k=1}^{K}\sum_{j=1}^{M_{\text{r}}}\mathbf{T}_{k}^{[j,j]}\Big[\beta\mu\Big(\mathbf{f}_{j,k}^{H}
(\xi\mathbf{I}_{M_{\text{t}}}+\mathbf{G}\mathbf{J}\mathbf{G}^{H})^{-1}\mathbf{G}\mathbf{J}\mathbf{G}^{H}\\ \nonumber
&(\xi\mathbf{I}_{M_{\text{t}}}+\mathbf{G}\mathbf{J}\mathbf{G}^{H})^{-1}
\mathbf{f}_{j,k}-2\mathbf{f}_{j,k}^{H}
(\mathbf{G}\mathbf{J}\mathbf{G}^{H}+\\  \nonumber
&\xi\mathbf{I}_{M_{\text{t}}})^{-1}\mathbf{f}_{j,k}\Big)+(1-\beta)(P_{\text{RF}}+2M_{\text{t}}P_{\text{A}})\Big] \\ \nonumber
=&\sum_{k=1}^{K}\sum_{j=1}^{M_{\text{r}}}\mathbf{T}_{k}^{[j,j]}\Bigg[(1-\beta)(P_{\text{RF}}+2M_{\text{t}}P_{\text{A}})-\\ \label{30}
&\qquad\qquad \beta\mu\sum_{i=1}^{M_{\text{t}}}
\frac{x_{i,j,k}(\mathbf{J}^{[i,i]}+2\xi)}{(\mathbf{J}^{[i,i]}+\xi)^{2}}\Bigg],
\end{align}
where we ignore the operation of $\Re(\cdot)$ as $\mathbf{f}_{j,k}^{H}
(\mathbf{G}\mathbf{J}\mathbf{G}^{H}+\xi\mathbf{I}_{M_{\text{t}}})^{-1}\mathbf{f}_{j,k}$ is a real value. From \eqref{30}, it can be observed that the contribution of vector $\mathbf{\bar{w}}_{j,k}$ to the objective can be expressed as $\mathcal{C}_{j,k}=-\sum_{i=1}^{M_{\text{t}}}
\frac{x_{i,j,k}(\mathbf{J}^{[i,i]}+2\xi)}{(\mathbf{J}^{[i,i]}+\xi)^{2}}$. Accordingly, with any given $\xi$, the optimal $\mathbf{T}_{k}$ can be obtained by
\begin{align}\label{31}
\mathbf{T}_{k}^{[j,j]}=\begin{cases}
&0, \quad \text{if}\quad \beta\mu\mathcal{C}_{j,k}+(1-\beta)\mathcal{\bar{C}} \geq 0,\\
&1 , \quad \text{if}\quad \beta\mu\mathcal{C}_{j,k}+(1-\beta)\mathcal{\bar{C}} < 0,
\end{cases} \quad \forall j, k,
\end{align}
where $\mathcal{\bar{C}}=(P_{\text{RF}}+2M_{\text{t}}P_{\text{A}})$. Based on the result in \eqref{31}, we can adopt the Bisection method to \eqref{29} to search for optimal $\xi$, which is summarized in \textbf{Algorithm \ref{Bisection}}.

In the second stage, the optimal hybrid beamformers can be obtained according to Proposition \ref{Proposition_1}.
\begin{algorithm}[t]
    \caption{The pseudo codes of Bisection algorithm.}
    \label{Bisection}
    \begin{algorithmic}[1]
        \STATE{Initialize initial $\xi_{\text{lower}}$ and $\xi_{\text{upper}}$. Set a convergence accuracy $\epsilon_{1}$.}
        \REPEAT
        \STATE{ $\xi=\frac{\xi_{\text{lower}}+\xi_{\text{upper}}}{2}$.}
        \STATE{ update $\{\mathbf{T}_{k}\}$ according to \eqref{31}.}
        \STATE{ update $\{\mathbf{\bar{w}}_{j,k}\}$ according to \eqref{29}.}
        \STATE{ \textbf{if} $\sum_{k=1}^{K}\sum_{j=1}^{M_{\text{r}}}\mathbf{T}_{k}^{[j,j]}\left(\sum_{i=1}^{M_{\text{t}}}
        \frac{x_{i,j,k}}{(\mathbf{J}^{[i,i]}+\xi)^{2}}\right)\geq P_{\text{max}}$}
        \STATE{  \quad $\xi_{\text{lower}}=\xi$,}
        \STATE{ \textbf{else}}
        \STATE{  \quad $\xi_{\text{upper}}=\xi$,}
        \STATE{ \textbf{end}}
        \UNTIL{ $\left|\sum_{k=1}^{K}\sum_{j=1}^{M_{\text{r}}}\mathbf{T}_{k}^{[j,j]}
        \left(\sum_{i=1}^{M_{\text{t}}}\frac{x_{i,j,k}}{(\mathbf{J}^{[i,i]}+\xi)^{2}}\right)- P_{\text{max}}\right|\leq\epsilon_{1}$.}
    \end{algorithmic}
\end{algorithm}

\subsection{Overall Algorithm}
The overall algorithm of the proposed WMMSE-TS algorithm is summarized in \textbf{Algorithm \ref{WMMSE-TS}}. In each iteration of the WMMSE-TS algorithm, the optimal or the stationary solutions are guaranteed for the corresponding subproblem, which indicates that the objective value remains non-decreasing over the iterations. Meanwhile, as the objective function is upper bounded due to the finite transmit power budget, the \textbf{Algorithm \ref{WMMSE-TS}} is guaranteed to converge to the stationary-point solution of the problem (P1).

The computational complexity of \textbf{Algorithm \ref{WMMSE-TS}} mainly arises from the matrix computation operations. To elaborate, the inverse matrix operations in steps 3 and 4 of \textbf{Algorithm \ref{WMMSE-TS}} incurs the complexity of $\mathcal{O}\big(M_{\text{r}}^{3}\big)$. The EVD operation in the step 5 of \textbf{Algorithm \ref{WMMSE-TS}} yields the complexity of $\mathcal{O}\big(M_{\text{t}}^{3}\big)$. For the QR decomposition in \eqref{30}, we consider employing the Householder transformation, which incurs the complexity of $\mathcal{O}\big(M_{\text{t}}T_{\text{s}}^{2}\big)$. Thus, the whole complexity of \textbf{Algorithm \ref{WMMSE-TS}} is given by $\mathcal{O}\Big(l_{\text{w}}(2M_{\text{r}}^{3}+M_{\text{t}}^{3}+\log_{2}(\frac{\xi_{\text{upper}}^{\text{in}}-
\xi_{\text{lower}}^{\text{in}}}{\epsilon_{1}}))+M_{\text{t}}T_{\text{s}}^{2}\Big)$, where $l_{\text{w}}$ denotes the number of WMMSE iterations, $\xi_{\text{lower}}^{\text{in}}$ and $\xi_{\text{upper}}^{\text{in}}$ denote the initial search lower and upper boundaries of the Bisection algorithm. Note that the proposed WMMSE-TS algorithm is still computationally efficient because the Bisection algorithm is carried out according to \eqref{29} instead of \eqref{28}, which avoids the high complexity involved in performing inverse matrix operation in each Bisection iteration.

\begin{algorithm}[t]
    \caption{The proposed WMMSE-TS algorithm.}
    \label{WMMSE-TS}
    \begin{algorithmic}[1]
        \STATE{Initialize $\{\mathbf{\Gamma}_{k},\mathbf{\bar{W}}_{k},\mathbf{T}_{k}\}$. Set a convergence accuracy $\epsilon_{2}$.}
        \REPEAT
        \STATE{ update $\mathbf{Z}_{k}$ according to \eqref{23}.}
        \STATE{ update $\mathbf{\Gamma}_{k}$ according to \eqref{25}.}
        \STATE{ Calculate the EVD results of $\sum_{i=1}^{K}\mathbf{H}_{i}^{H}\mathbf{Z}_{i}\mathbf{\Gamma}_{i} \mathbf{Z}_{i}^{H}\mathbf{H}_{i}$.}
        \STATE{ update $\mathbf{\bar{W}}_{k}$ and $\mathbf{T}_{k}$ by carrying out \textbf{Algorithm \ref{Bisection}}.}
        \UNTIL{ the objective value converges with $\epsilon_{2}$.}
        \STATE{ update $\{\mathbf{P}_{1},\mathbf{P}_{2},\mathbf{P},\mathbf{W}_{\text{BB}}\}$ according to Proposition \ref{Proposition_1}.}
    \end{algorithmic}
\end{algorithm}
\section{Hybrid Beamforming Design with Discrete Phase Shifters}\label{Section_4}
In this section, we consider the hybrid beamforming optimization under the practical discrete phase shifters. By rendering the equality constraint to be the penalty terms of the objective function, a PLI optimization framework is proposed. In particular, the optimal analog and baseband digital beamformers for each iteration are derived in the closed-form expressions.

\subsection{Problem Reformulation}
To facilitate the analog beamforming optimization, we introduce the fully-digital beamformer $\mathbf{\bar{W}}_{k}$ to replace $\mathbf{P}\mathbf{W}_{k}$ and move the equality constraint $\mathbf{\bar{W}}_{k}\mathbf{T}_{k}-\mathbf{P}\mathbf{W}_{k}$ to the penalty of the objective function. With the WMMSE transformation in \eqref{18}, the optimization problem (P1) is reformulated as
\begin{subequations}
\begin{align}
\nonumber (\text{P}3.1) \max\limits_{\mathbf{\bar{W}}_{k},\mathbf{\Gamma}_{k},\mathbf{Z}_{k},
\atop
\mathbf{T}_{k},\mathbf{P},\mathbf{W}_{k}}& \sum_{k=1}^{K}
\beta\mu\Big(\log_{2}|\mathbf{\Gamma}_{k}|\!-\!\text{Tr}(\mathbf{\Gamma}_{k}\mathbb{F}(\mathbf{Z}_{k},\mathbf{\bar{W}}_{k},\\  \nonumber
&\mathbf{T}_{k}))+M_{\text{r}}\!-\!\frac{1}{2\varrho}\|\mathbf{\bar{W}}_{k}\mathbf{T}_{k}-\mathbf{P}\mathbf{W}_{k}\|_{F}^{2}\Big)\\ \label{32a}
&-(1-\beta)\text{Tr}(\mathbf{T}_{k})(P_{\text{RF}}+M_{\text{t}}P_{\text{A}})\\
\label{32b}\text{s.t.} \quad & \eqref{19b},\eqref{19c},
\end{align}
\end{subequations}
where $\varrho$ denotes the scaling factor of the penalty terms. However, the objective function cannot be directly tackled as in the previous section due to the existence of $\mathbf{T}_{k}$. To tackle this issue, we construct a conservative upper bound of the penalty \cite{Matrix}, which is given by
\begin{align}\label{33}
\|\mathbf{\bar{W}}_{k}\mathbf{T}_{k}-\mathbf{P}\mathbf{W}_{k}\|_{F}^{2}\leq \sum_{j=1}^{M_{\text{r}}}\mathbf{T}_{k}^{j,j}\|\mathbf{\bar{w}}_{j,k}-\mathbf{P}\mathbf{w}_{j,k}\|^{2}.
\end{align}
Here, $\mathbf{w}_{j,k}$ denotes the $j$-th column vector of $\mathbf{W}_{k}$. Note that according to the property of matrix norm, it readily knows that the right-half side of \eqref{33} is the tight approximation of $\|\mathbf{\bar{W}}_{k}\mathbf{T}_{k}-\mathbf{P}\mathbf{W}_{k}\|_{F}^{2}$ when $\|\mathbf{\bar{W}}_{k}\mathbf{T}_{k}-\mathbf{P}\mathbf{W}_{k}\|_{F}^{2}\rightarrow 0$. Accordingly, we reformulate the problem (P1) as
\begin{subequations}
\begin{align}
\nonumber (\text{P}3.2) \max\limits_{\mathbf{\bar{W}}_{k},\mathbf{\Gamma}_{k},\mathbf{Z}_{k},
\atop
\mathbf{T}_{k},\mathbf{P},\mathbf{W}_{k}} & \sum_{k=1}^{K}
\beta\mu\Big(\log_{2}|\mathbf{\Gamma}_{k}|\!-\!\text{Tr}(\mathbf{\Gamma}_{k}\mathbb{F}(\mathbf{Z}_{k},\mathbf{\bar{W}}_{k},\\ \nonumber
&\!\!\!\!\!\mathbf{T}_{k}))+M_{\text{r}}-\frac{1}{2\varrho}\sum_{j=1}^{M_{\text{r}}}\mathbf{T}_{k}^{j,j}
\|\mathbf{\bar{w}}_{j,k}-\mathbf{P}\mathbf{w}_{j,k}\|^{2}\Big)\\ \label{34a}
&\!\!\!\!\!-(1-\beta)\text{Tr}(\mathbf{T}_{k})(P_{\text{RF}}+M_{\text{t}}P_{\text{A}})\\
\label{34b}\text{s.t.} \quad & \eqref{19b},\eqref{19c}.
\end{align}
\end{subequations}
Similarly, since all the variables are separated in the constraints, we adopt the layered BCD optimization method to solve the problem (P3.2). In the outer layer, we update $\varrho$ to ensure the feasibility of the optimized fully-digital beamformer. While in the inner layer, we iteratively optimize $\{\mathbf{\bar{W}}_{k},\mathbf{T}_{k},\mathbf{\Gamma}_{k},\mathbf{Z}_{k},\mathbf{P},\mathbf{W}_{k}\}$ in different blocks.

\subsection{Proposed PLI Algorithm}

\subsubsection{Subproblem with respect to $\{\mathbf{Z}_{k},\mathbf{\Gamma}_{k}\}$}
Recall the optimization process in the previous section, the optimal $\{\mathbf{Z}_{k},\mathbf{\Gamma}_{k}\}$ are derived by
\begin{align}\label{35}
\mathbf{Z}_{k} &= \left(\mathbf{\dot{Q}}_{k}+\mathbf{H}_{k}\mathbf{\bar{W}}_{k}\mathbf{T}_{k}\mathbf{T}_{k}^{H}
\mathbf{\bar{W}}_{k}^{H}\mathbf{H}_{k}^{H}\right)^{-1}\mathbf{H}_{k}\mathbf{\bar{W}}_{k}\mathbf{T}_{k},\\ \label{36}
\mathbf{\Gamma}_{k} &= \mu^{-1}\mathbb{F}(\mathbf{Z}_{k},\mathbf{\bar{W}}_{k},\mathbf{T}_{k}))^{-1}.
\end{align}

\subsubsection{Subproblem with respect to $\{\mathbf{\bar{W}}_{k},\mathbf{T}_{k}\}$} the optimization problem with respect to $\{\mathbf{\bar{W}}_{k},\mathbf{T}_{k}\}$ is transformed into
\begin{subequations}
\begin{align}\nonumber
 (\text{P}3.3)\  \min\limits_{\mathbf{\bar{W}}_{k},\mathbf{T}_{k}}\  & \sum_{k=1}^{K}
\beta\mu\Big(\text{Tr}(\mathbf{\Gamma}_{k}\mathbb{F}(\mathbf{Z}_{k},\mathbf{\bar{W}}_{k},\mathbf{T}_{k}))+\\ \nonumber
&\frac{1}{2\varrho}\sum_{j=1}^{M_{\text{r}}}\mathbf{T}_{k}^{j,j}\|\mathbf{\bar{w}}_{j,k}-\mathbf{P}\mathbf{w}_{j,k}\|^{2}\Big)+\\ \label{37a}
&(1-\beta)(P_{\text{RF}}+M_{\text{t}}P_{\text{A}})\text{Tr}(\mathbf{T}_{k})\\
\label{37b}\text{s.t.} \quad & \eqref{19b},\eqref{19c}.
\end{align}
\end{subequations}
Nevertheless, the problem (P3.3) is still intractable to handle due to coupling between $\mathbf{\bar{W}}_{k}$ and $\mathbf{T}_{k}$. To deal with this issue, we rewrite the problem (P3.3) as
\begin{subequations}
\begin{align}\nonumber
 (\text{P}3.4)\  \min\limits_{\mathbf{\bar{w}}_{j,k},\mathbf{T}_{k}}\  &\sum_{k=1}^{K}\sum_{j=1}^{M_{\text{r}}}\mathbf{T}_{k}^{[j,j]}\Big[\beta\mu
 \Big(\mathbf{\bar{w}}_{j,k}^{H}\Big(
\frac{1}{2\varrho}\mathbf{I}_{M_{\text{t}}}+\\ \nonumber
&\!\!\!\!\!\!\!\!
\sum\nolimits_{i=1}^{K}\mathbf{H}_{i}^{H}\mathbf{Z}_{i}\mathbf{\Gamma}_{i} \mathbf{Z}_{i}^{H}\mathbf{H}_{i}\Big)
\mathbf{\bar{w}}_{j,k}-2\Re\Big(\mathbf{\bar{w}}_{j,k}^{H}\mathbf{m}_{j,k}\Big)\Big)\\ \label{38a}
&\!\!\!\!\!\!\!\!+(1-\beta)(P_{\text{RF}}+M_{\text{t}}P_{\text{A}})\Big]\\
\label{38b}\text{s.t.} \quad & \sum_{k=1}^{K}\sum_{j=1}^{M_{\text{r}}}\mathbf{T}_{k}^{[j,j]}\|\mathbf{\bar{w}}_{j,k}\|^{2}_{\text{F}}\leq P_{\text{max}},,\\
\label{38c} & \eqref{19c},
\end{align}
\end{subequations}
where $\mathbf{M}_{k}=\mathbf{H}_{k}^{H}\mathbf{Z}_{k}\mathbf{\Gamma}_{k}+\frac{1}{2\varrho}\mathbf{P}\mathbf{W}_{k}=[\mathbf{m}_{1,k},\cdots, \mathbf{m}_{M_{\text{r}},k}]$. According to the first-order optimality condition, the optimal $\mathbf{\bar{w}}_{j,k}$ can derived by
\begin{align}\label{39}
\mathbf{\bar{w}}_{j,k} =\left(\frac{1+2\varrho\xi}{2\varrho}\mathbf{I}_{M_{\text{t}}}+\sum_{i=1}^{K}\mathbf{H}_{i}^{H}\mathbf{Z}_{i}\mathbf{\Gamma}_{i} \mathbf{Z}_{i}^{H}\mathbf{H}_{i}\right)^{-1}\!\!\!\mathbf{m}_{j,k}.
\end{align}
Then, by substituting \eqref{39} into the objective function \eqref{38a}, we can obtain
\begin{align}\nonumber
\eqref{38a} \!=\! &\sum_{k=1}^{K}\sum_{j=1}^{M_{\text{r}}}\mathbf{T}_{k}^{[j,j]}\Bigg[\beta\mu\Bigg(\mathbf{m}_{j,k}^{H}
\left(\frac{1+2\varrho\xi}{2\varrho}\mathbf{I}_{M_{\text{t}}}\!+\!\mathbf{G}\mathbf{J}\mathbf{G}^{H}\right)^{-1}\\ \nonumber
&\mathbf{G}\mathbf{J}\mathbf{G}^{H}\left(\frac{1+2\varrho\xi}{2\varrho}\mathbf{I}_{M_{\text{t}}}
+\mathbf{G}\mathbf{J}\mathbf{G}^{H}\right)^{-1}
\mathbf{m}_{j,k}-\\ \nonumber
&2\mathbf{m}_{j,k}^{H}
\left(\frac{1+2\varrho\xi}{2\varrho}\mathbf{I}_{M_{\text{t}}}+\mathbf{G}\mathbf{J}\mathbf{G}^{H}\right)^{-1}\mathbf{m}_{j,k}\Bigg)+ \\ \nonumber
&(1-\beta)(P_{\text{RF}}+M_{\text{t}}P_{\text{A}})\Bigg] \\ \nonumber
=&\sum_{k=1}^{K}\sum_{j=1}^{M_{\text{r}}}\mathbf{T}_{k}^{[j,j]}\Bigg[(1-\beta)(P_{\text{RF}}+M_{\text{t}}P_{\text{A}})-\\ \label{40}
&\qquad\qquad \beta\mu\sum_{i=1}^{M_{\text{t}}}
\frac{y_{i,j,k}\left(\mathbf{J}^{[i,i]}+\frac{1+2\varrho\xi}{\varrho}\right)}
{\left(\mathbf{J}^{[i,i]}+\frac{1+2\varrho\xi}{2\varrho}\right)^{2}}\Bigg].
\end{align}
where $y_{i,j,k}=(\mathbf{G}^{H}\mathbf{m}_{j,k}\mathbf{m}_{j,k}^{H}\mathbf{G})^{[i,i]}$. Consequently, given any feasible Lagrangian
multiplier $\xi$, the optimal $\mathbf{T}_{k}$ can be obtained by
\begin{align}\label{41}
\mathbf{T}_{k}^{[j,j]}=\begin{cases}
&0, \quad \text{if}\quad \beta\mu\mathcal{E}_{j,k}+(1-\beta)\mathcal{\bar{E}} \geq 0,\\
&1 , \quad \text{if}\quad \beta\mu\mathcal{E}_{j,k}+(1-\beta)\mathcal{\bar{E}} < 0,
\end{cases} \quad \forall j, k,
\end{align}
where $\mathcal{E}_{j,k}=\sum_{i=1}^{M_{\text{t}}}
\frac{y_{i,j,k}\left(\mathbf{J}^{[i,i]}+\frac{1+2\varrho\xi}{\varrho}\right)}{\left(\mathbf{J}^{[i,i]}+\frac{1
+2\varrho\xi}{2\varrho}\right)^{2}}$ and $\mathcal{\bar{E}}=\\ \frac{y_{i,j,k}\left(\mathbf{J}^{[i,i]}+\frac{1+2\varrho\xi}{\varrho}\right)}
{\left(\mathbf{J}^{[i,i]}+\frac{1+2\varrho\xi}{2\varrho}\right)^{2}}$. As a result, $\{\mathbf{\bar{W}}_{k},\mathbf{T}_{k}\}$ can be jointly optimized by employing the Bisection approach to the following equality
\begin{align}\label{42}
\sum_{k=1}^{K}\sum_{j=1}^{M_{\text{r}}}\mathbf{T}_{k}^{[j,j]}\left(\sum_{i=1}^{M_{\text{t}}}\frac{y_{i,j,k}}{
\left(\mathbf{J}^{[i,i]}+\frac{1+2\varrho\xi}{2\varrho}\right)^{2}}\right)
=P_{\text{max}}.
\end{align}

\subsubsection{Subproblem with respect to $\{\mathbf{P}\}$} With the optimized $\mathbf{T}_{k}$, the number of activated RF chains are updated by $T_{\text{s}}=\sum_{k=1}^{K}\mathbf{T}_{k}$. Thus, the analog beamformer optimization subproblem can be formulated as
\begin{subequations}
\begin{align}
\label{43a} (\text{P}3.5)\  \min\limits_{\mathbf{P}}\  &\sum_{k=1}^{K}\sum_{j=1}^{M_{\text{r}}}\mathbf{T}_{k}^{j,j}\|\mathbf{\bar{w}}_{j,k}-\mathbf{P}
\mathbf{\tilde{w}}_{j,k,T_{\text{s}}}\|^{2}\\
\label{43b}\text{s.t.} \quad &  \mathbf{P} = \mathbf{P}_{1}+\mathbf{P}_{2},\\
\label{43c}&\mathbf{P}_{1},\mathbf{P}_{2}\in \Phi_{2},
\end{align}
\end{subequations}
where $\mathbf{\tilde{w}}_{j,k,T_{\text{s}}}$ is a subvector of $\mathbf{w}_{j,k}$ consisting of the former $T_{\text{s}}$ elements of $\mathbf{w}_{j,k}$. Due to the constraint \eqref{43c}, the problem (P3.5) cannot be directly optimized. However, it is easy to find that the objective function is convex over the continuous phase-shift region, so the optimal continuous phase shifters can be obtained by ignoring the discrete phase-shift constraints and solving the problem (P3.5). Let $\mathbf{P}_{\text{c}}$ denote the optimized continuous phase shifter through the standard convex toolbox (e.g., CVX), the $h$-th row and the $v$-column element of the discrete phase shifters can be obtained by
\begin{align}
\label{44} \mathbf{P}^{[h,v]}=\text{arg} \min\limits_{\mathbf{P}_{\text{d}}^{[h,v]}\in\Phi_{3}}\left|\mathbf{P}_{\text{d}}^{[h,v]}-\mathbf{P}_{\text{c}}^{[h,v]}\right|,
\end{align}
where $\Phi_{3}=\{\mathbf{P}|\mathbf{P}=\mathbf{P}_{1}+\mathbf{P}_{2},\mathbf{P}_{1},\mathbf{P}_{1}\in\Phi_{2}\}$ denotes the feasible region of the discrete phase shifters.

\subsubsection{Subproblem with respect to $\{\mathbf{W}_{k}\}$} The subproblem for baseband beamformer optimization can formulated as
\begin{subequations}
\begin{align}
\label{45a} (\text{P}3.6)\  \min\limits_{\mathbf{e}_{j,k}}\  &\sum_{k=1}^{K}\sum_{j=1}^{M_{\text{r}}}\mathbf{T}_{k}^{j,j}\|\mathbf{P}\mathbf{w}_{j,k}-\mathbf{\bar{w}}_{j,k}\|^{2}
\end{align}
\end{subequations}
The problem (P3.7) is a standard unconstrained convex program, where the optimal $\mathbf{e}_{j,k}$ can be derived by
\begin{align}\label{46}
\mathbf{w}_{j,k} =\begin{cases}
&\mathbf{P}^{\dag}\mathbf{\bar{w}}_{j,k}, \quad \text{if}\quad \mathbf{T}_{k}^{j,j} = 1,\\
&\mathbf{0} , \qquad \quad \,\  \text{if}\quad \mathbf{T}_{k}^{j,j} = 0,
\end{cases} \quad \forall j, k.
\end{align}

\subsubsection{Outer layer update}
In the outer layer of the PLI algorithm, the penalty factor $\varrho$ decreases with a constant coefficient $\bar{c}<1$, i.e.,
\begin{align}\label{47}
\varrho = \bar{c}\varrho.
\end{align}

\subsection{Overall Algorithm}
The proposed PLI algorithm is summarized in \textbf{Algorithm \ref{PLI}}. In the inner layer iterations, each subproblem can be optimally solved under the fixed scaling factor $\varrho$, which indicates that the objective value remains non-decreasing over the iterations. In the outer layer iterations, by steadily decreasing the scaling factor $\varrho$, the PLI algorithm can converge to at least the stationary-point solutions that satisfied $\|\mathbf{\bar{W}} -\mathbf{P}\mathbf{W}_{\text{BB}}\|_{F}\rightarrow0$  \cite{PDD}. Let $l_{\text{inn}}$ and $l_{\text{out}}$ denote the number of inner and outer layer iterations of \textbf{Algorithm \ref{PLI}}. The overall computational complexity of the proposed PLI algorithm is given by $\mathcal{O}\Big(l_{\text{out}}l_{\text{inn}}(2M_{\text{r}}^{3}+M_{\text{t}}^{3}+\log_{2}(\frac{\xi_{\text{upper}}^{\text{in}}-
\xi_{\text{lower}}^{\text{in}}}{\epsilon_{1}})+T_{\text{s}}^{3})\Big)$.

\begin{algorithm}[t]
    \caption{PIL algorithm.}
    \label{PLI}
    \begin{algorithmic}[1]
        \STATE{Initialize $\{\mathbf{\Gamma}_{k},\mathbf{\bar{W}}_{k},\mathbf{T}_{k},\mathbf{P},\mathbf{W}_{k}\}$. Set the convergence accuracy $\epsilon_{3}$ and a tolerant threshold $\epsilon_{4}$.}
        \REPEAT
        \REPEAT
        \STATE{ update $\mathbf{Z}_{k}$ according to \eqref{35}.}
        \STATE{ update $\mathbf{\Gamma}_{k}$ according to \eqref{36}.}
        \STATE{ update $\mathbf{\bar{W}}_{k}$ and $\mathbf{T}_{k}$ by carrying out the Bisection algorithm.}
        \STATE{ update $\mathbf{P}$ according to \eqref{44}.}
        \STATE{ update $\mathbf{W}_{k}$ according to \eqref{46}.}
        \UNTIL{the objective value converges with $\epsilon_{3}$.}
        \STATE{ update $\varrho$ according to \eqref{47}.}
        \UNTIL{ the penalty term falls below the tolerant threshold $\epsilon_{4}$.}
    \end{algorithmic}
\end{algorithm}

\section{Numerical Results}\label{Section_5}
\begin{figure}[t]
  \centering
  \includegraphics[scale = 0.8]{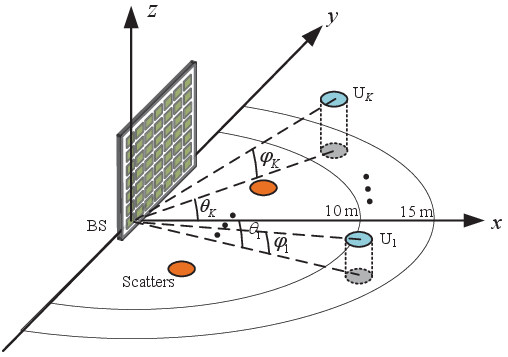}
  \caption{Simulation setup of the considered near-field network.}
  \label{Fig.3}
\end{figure}

In this section, the numerical simulation results are provided to demonstrate the effectiveness of the proposed scheme. The location topology setup is illustrated in Fig. \ref{Fig.3}, where the BS is located at coordinates (0,0,0) meter (m). The multi-antenna users are randomly distributed in a circular ring of 5 m in width centered on the BS, where the azimuth and elevation angles of each user randomly take values in the interval $[-\frac{\pi}{2},\frac{\pi}{2}]$. While in the circle area with a radius of $10$ m, $5$ scatters are positioned with random azimuth and elevation angles. The power consumption of each RF chain is set as $0.2$ watt (W), and the power consumption of each unit-modulus phase shifter is $0.01$ W \cite{PS_power_consumption}. Unless otherwise specified, the simulation parameters are generally set as Table \ref{tab-1}. Each result is averaged over $100$ Monte Carle experiments.

For performance comparison, we consider three baseline schemes in the numerical results:
\begin{itemize}
  \item \textbf{Far-field channel model}: In this baseline scheme, the planar-wave based far-field channel model is used to characterize the signal propagation in the considered network. Specifically, the far-field channels between the BS and $\text{U}_{k}$ are modeled by the product of array response vectors, i.e., $\mathbf{H}_{k} = \tilde{\beta}_{k}\mathbf{a}_{\text{FF}}^{k}(r_{k},\theta_{k},\varphi_{k})(\mathbf{a}_{\text{FF}}^{\text{BS}}
(r_{k},\theta_{k},\varphi_{k}))^{T}+\sum_{c=1}^{C}\tilde{\beta}_{c}\mathbf{a}_{\text{FF}}^{k}(r_{c},\theta_{c},\varphi_{c})(\mathbf{a}_{\text{FF}}^{\text{BS}}
(r_{c},\theta_{c},\varphi_{c}))^{T}$, where the $\mathbf{a}_{\text{FF}}^{k}$ and $\mathbf{a}_{\text{FF}}^{\text{BS}}$ denote the corresponding array response vectors at the $\text{U}_{k}$ and BS.
  \item \textbf{Fixed data-stream allocation}: In the fixed data-stream allocation scheme, the BS activates $K\bar{t}$ RF chains and transmits $\bar{t}$ data streams to each user. Since the HPC at the BS is not considered in this scheme, we directly adopt the existing PDD algorithm \cite{PDD} to maximize the achievable rate of multiple users.
  \item \textbf{Conventional hybrid beamforming}: In the conventional hybrid beamforming scheme, each RF is connected to an antenna only via one phase shifter. To achieve the same performance as fully-digital beamforming for fairness comparison, we activated the RF chains twice as many as the amount of the sent data streams, where the optimal analog and baseband digital beamformers can be obtained according to \cite[Proposition 2]{Millimeter2}.
\end{itemize}

\begin{table}[t]
\centering
    \caption{System parameters}\label{tab-1}
\begin{tabular}{|c|c|}
		\hline
        \tabincell{c}{Operating carrier frequency} & $f = 28$ GHz \\
		\hline
        \tabincell{c}{Number of transmit/receive antennas} & $M_{\text{t}} = 512$, $M_{\text{r}} = 10$\\
		\hline
        \tabincell{c}{Number of equipped  RF chains \\ at the BS} & $M_{\text{t}}^{\text{RF}} = 40$\\
		\hline
        \tabincell{c}{Antenna space} & $d = \frac{\lambda}{2}=\frac{c}{2f}$ m \\
		\hline
		\tabincell{c}{Noise power at receivers} & $\sigma^{2}=-105$ dBm \\
		\hline
        \tabincell{c}{Initial Bisection boundary} & $\xi_{\text{lower}}=0$, $\xi_{\text{upper}}=10^{8}$ \\
		\hline
        \tabincell{c}{Constant scaling coefficients} & $\mu=1.5$, $\bar{c}=0.75$\\
        \hline
        \tabincell{c}{Weight coefficient} & $\beta = 0.7$\\
		\hline
        \tabincell{c}{Number of users} & $K = 4$\\
		\hline
        \tabincell{c}{Transmit/Receive antenna \\ gains at the BS} &  $G_{\text{t}}=G_{\text{r}}=0$ dB\\
        \hline \tabincell{c}{convergence accuracy}
         & \tabincell{c}{$\epsilon_{1}=10^{-6}$, \\ $\epsilon_{2}=\epsilon_{3}=\epsilon_{4}=10^{-2}$} \\
		\hline
\end{tabular}
\end{table}

\begin{figure}[t]
  \centering
  \includegraphics[scale = 0.52]{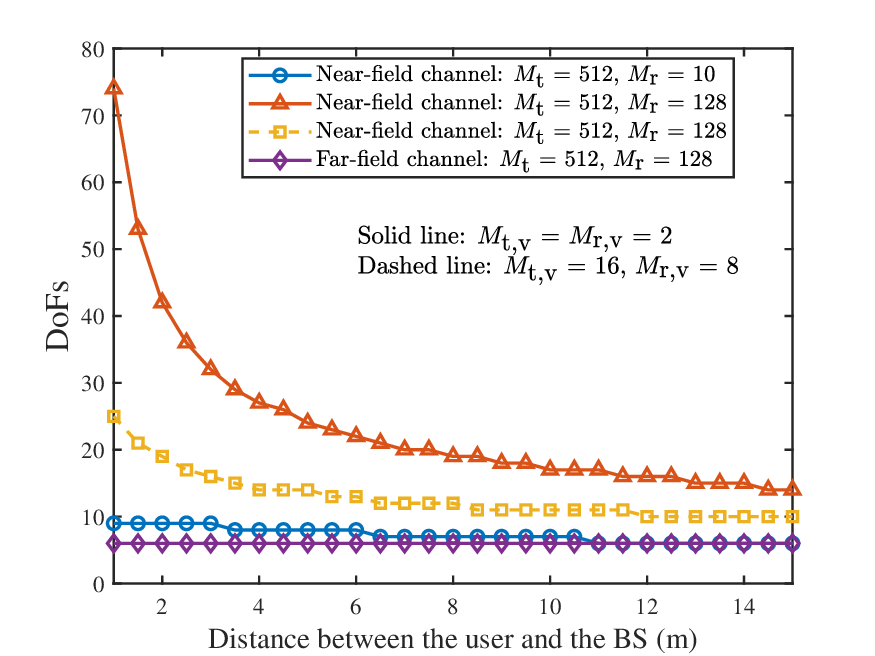}
  \caption{Spatial EDoF of near-field/far-field MIMO channel model versus the distance between the receiver and the BS.}
  \label{Fig.4}
\end{figure}

\begin{figure}[t]
 \centering
 \subfigure[The achievable rate versus the distance between the BS and the user for different hybrid beamforming architectures.]{
  \includegraphics[scale = 0.52]{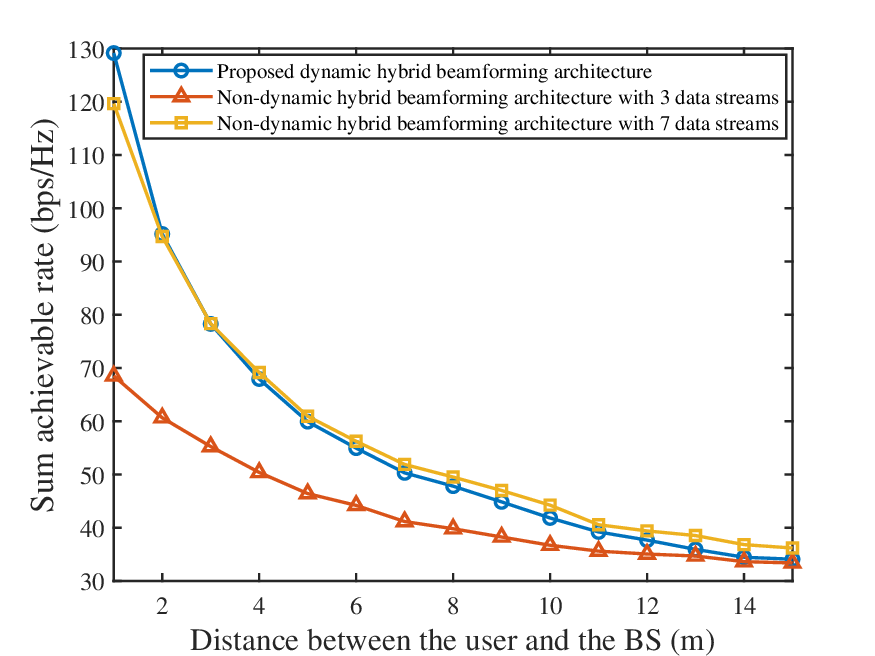}
   \label{Fig.5a}}
\subfigure[The HPC versus the distance between the BS and the user for different hybrid beamforming architectures.]{
 \includegraphics[scale = 0.52]{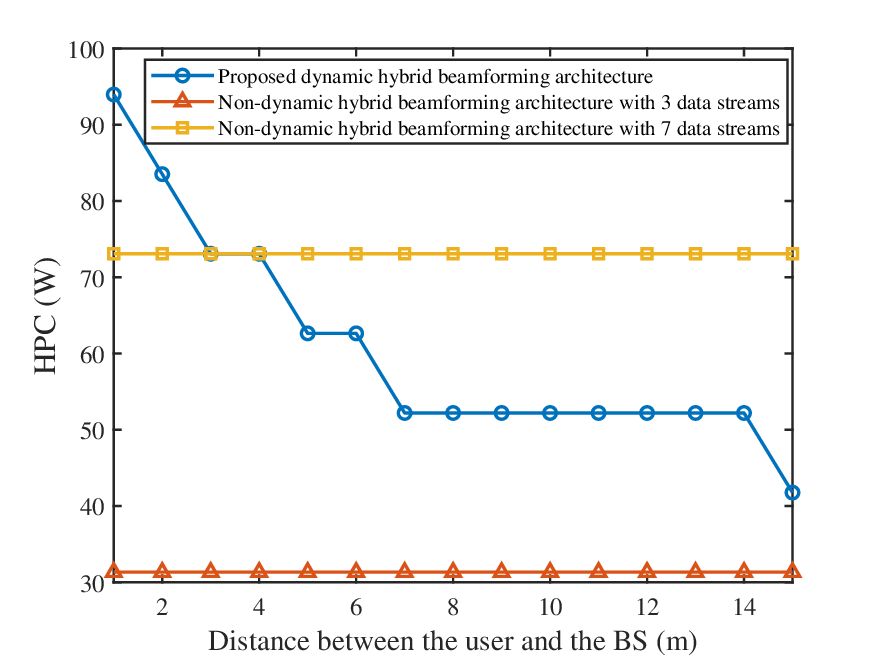}
 \label{Fig.5b}}
 \caption{Performance comparison of different hybrid beamforming architectures.}
\label{Fig.5}
\end{figure}

In Fig. \ref{Fig.4}, the effective DoF (EDoF) of the near-field/far-field MIMO channel model versus the distance between the BS and the receiver under different antenna setups is presented. Note that we focus on the EDoF metric as it represents the number of available independent sub-channels for communications, which is calculated by the dominant singular values of the channel matrix \cite{NF_EDoF_mag}. It can be observed from Fig. \ref{Fig.4} that the near-field channel introduces more spatial EDoF than the far-field channel. This can be expected because, from the physical signal propagation perspective, the near-field array responses at different antennas orientated towards the same reference point are not parallel, which degrades the relevance degree between array response vectors, thus resulting in an increased rank of the channel matrix. From Fig. \ref{Fig.4}, it is also shown that an upward trend in the number of receiving antennas has a positive impact on the increase of EDoF, which is due to the fact that increasing receiving antennas introduces more signal propagation links and favors an expansion of the number of effective sub-channels. Furthermore, we can observe that the antenna topologies at the transceivers also affect the EDoF of near-field channels. To elaborate, the smaller $M_{\text{t},\text{v}}$ and $M_{\text{t},\text{h}}$ lead to the larger antenna aperture, which thus expands the near-field region and brings more spatial EDoF.

The communication rate and HPC achieved by different hybrid beamforming architectures are depicted in Fig. \ref{Fig.5}, where the switch module is dropped in the non-dynamic hybrid beamforming architecture. Note that a single-user scenario is considered in Fig. \ref{Fig.5}. It can be observed from Fig. \ref{Fig.5a} that the proposed dynamic hybrid beamforming architecture consistently outperforms the 3-data-stream non-dynamic hybrid beamforming architecture in terms of communication performance. This is because the non-dynamic hybrid beamforming architecture cannot fully utilize the spatial DoF of near-field channels, whereas the dynamic hybrid beamforming architecture allows activating more RF chains for spatial multiplexing gain enhancement. It is also observed from Fig. \ref{Fig.5b} that when the transmission distance is larger than 4 m, the proposed dynamic hybrid beamforming architecture consumes less HPC than the non-dynamic hybrid beamforming architecture with 7 data streams. This can be expected as the spatial DoF of the considered near-field channel falls below 7 as the distance grows, and hence the dynamic hybrid beamforming architecture adaptively turns off the extra RF chains to save HPC.

\begin{figure}[t]
  \centering
  \includegraphics[scale = 0.52]{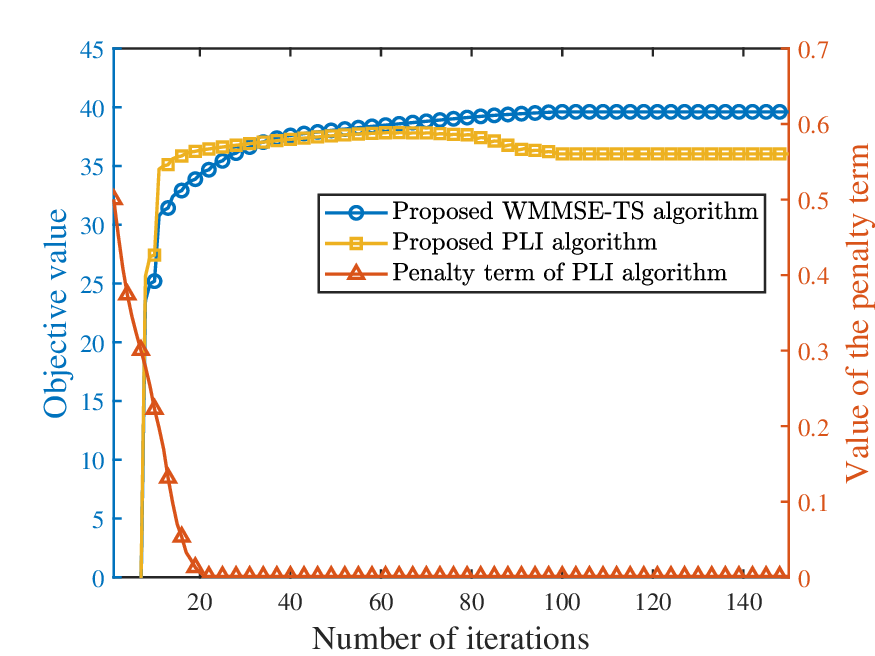}
  \caption{Convergence performance of the proposed algorithms with $P_{\text{max}}$ = 15 dBm.}
  \label{Fig.6}
\end{figure}

The convergence performance of the proposed algorithms is illustrated in Fig. \ref{Fig.6}, where the 3-bit discrete phase shifters are considered for the PLI algorithm. As can be observed from Fig. \ref{Fig.6}, with the increase of the number of iterations, the objective values of \textbf{Algorithm \ref{WMMSE-TS}} and \textbf{Algorithm \ref{PLI}} can converge to the stationary-point solutions within the finite steps, and meanwhile, the penalty term $\sum_{j=1}^{M_{\text{r}}}\mathbf{T}_{k}^{j,j}\|\mathbf{\bar{w}}_{j,k}-\mathbf{P}\mathbf{w}_{j,k}\|^{2}$ of the PLI algorithm approaches $0$ in around 20 iterations, which verify the effectiveness of our designed algorithms. It can be also found that the objective value of the PLI algorithm experiences an upward, followed by a downward, and then a stabilizing trend. This can be explained as follows. In the initial stage of the algorithm, we generally use a relatively large $\varrho$ scaling factor to render it easy to find the feasible solution to the optimization problem, so the objective value increases over the BCD iterations. After several iterations, $\varrho$ becomes small and the PLI algorithm is forced to focus on squeezing the gap between the hybrid beamforming and fully-digital beamforming, which however, sacrifices the communication/HPC performance of the considered network. Then, when $\sum_{j=1}^{M_{\text{r}}}\mathbf{T}_{k}^{j,j}\|\mathbf{\bar{w}}_{j,k}-\mathbf{P}\mathbf{w}_{j,k}\|^{2}\rightarrow 0$ is satisfied, the penalty term becomes the dispensable part of the objective function, and the objective value converges over the BCD iterations.

\begin{figure}[t]
 \centering
 \subfigure[The objective value versus the transmit power at the BS for different schemes.]{
  \includegraphics[scale = 0.52]{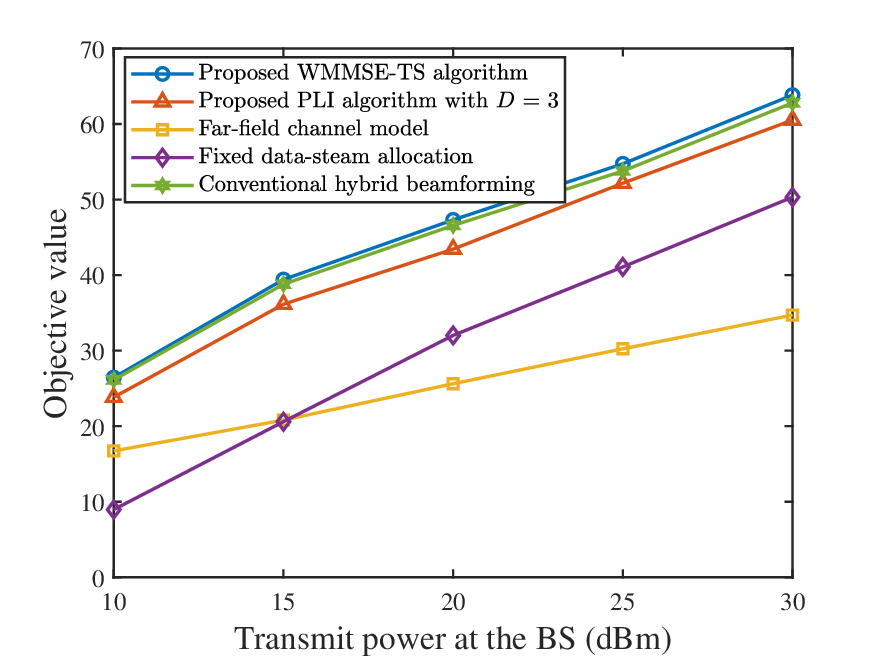}
   \label{Fig.7a}}
\subfigure[The achievable rate and hardware power consumption versus the transmit power at the BS for different schemes.]{
 \includegraphics[scale = 0.52]{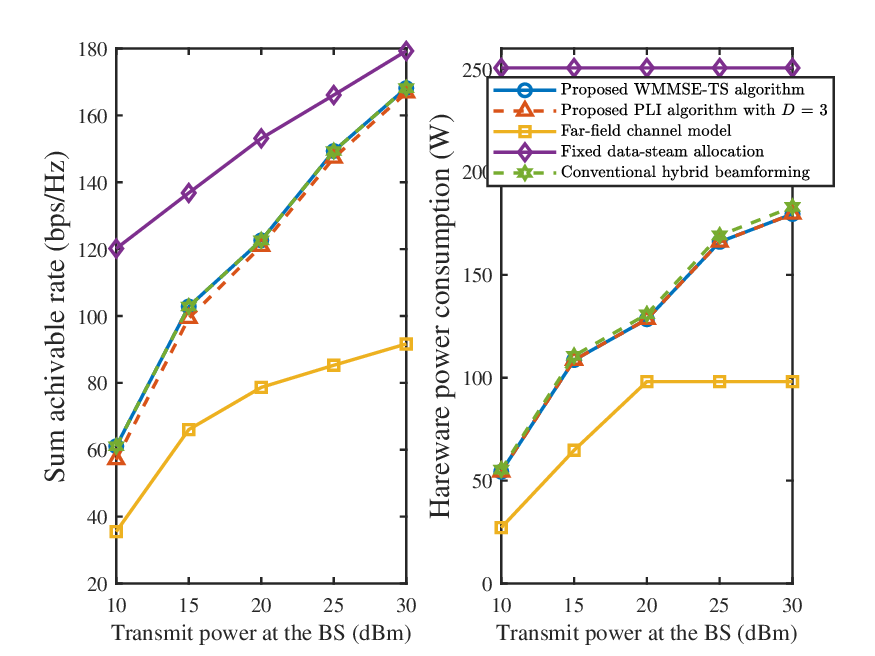}
 \label{Fig.7b}
}\caption{Performance comparison with different baseline schemes with $\bar{t}=8$.}
\label{Fig.7}
\end{figure}

Fig. \ref{Fig.7} compares the performance of the proposed scheme with that of the other baseline schemes, where the objective value achieved by different schemes are illustrated in Fig. \ref{Fig.7a} and their corresponding achievable rate and HPC are present in Fig. \ref{Fig.7b}. From Fig. \ref{Fig.7a}, it can be observed that the WMMSE-TS algorithm realizes a higher objective value than the PLI algorithm. This is because the consideration of discrete phase shifters inevitably causes quantification errors for the analog beamformer optimization, which deteriorates the communication performance of the network. Also, we can observe that the near-field channel model outperforms the far-field channel model with respect to the objective value. This is because although using the far-field channel model consumes low HPC, the near-field channel model provides larger multiplexing gains than the far-field channel model, which serves to be the dominant factor in the objective function under the given weight coefficient $\beta$. Moreover, Fig. \ref{Fig.7} shows that the proposed WMMSE-TS algorithm achieves the same communication performance as the conventional hybrid beamforming but consumes less hardware power. This is because both dynamic and conventional hybrid beamforming can achieve the optimal performance of fully-digital beamforming, whereas the conventional hybrid beamforming requires more RF chains than the dynamic hybrid beamforming architecture. For the fixed data-stream allocation scheme, as we allow the BS to send more data streams, it achieves the highest communication rate compared to the other schemes. Nevertheless, the fixed data-stream allocation scheme also suffers the highest HPC, resulting in the lowest objective value among all the schemes.

\begin{figure}[t]
 \centering
 \subfigure[The achievable rate versus the hardware power consumption under different $P_{\text{max}}$.]{
  \includegraphics[scale = 0.52]{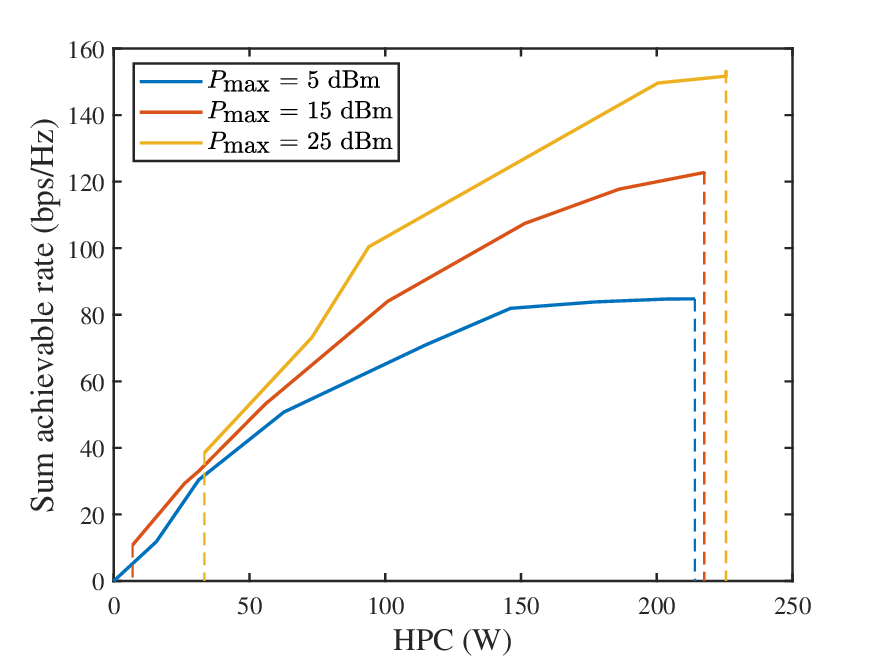}
   \label{Fig.8a}}
\subfigure[The achievable rate and hardware power consumption versus the weight factor $\beta$ under different $\mu$.]{
 \includegraphics[scale = 0.52]{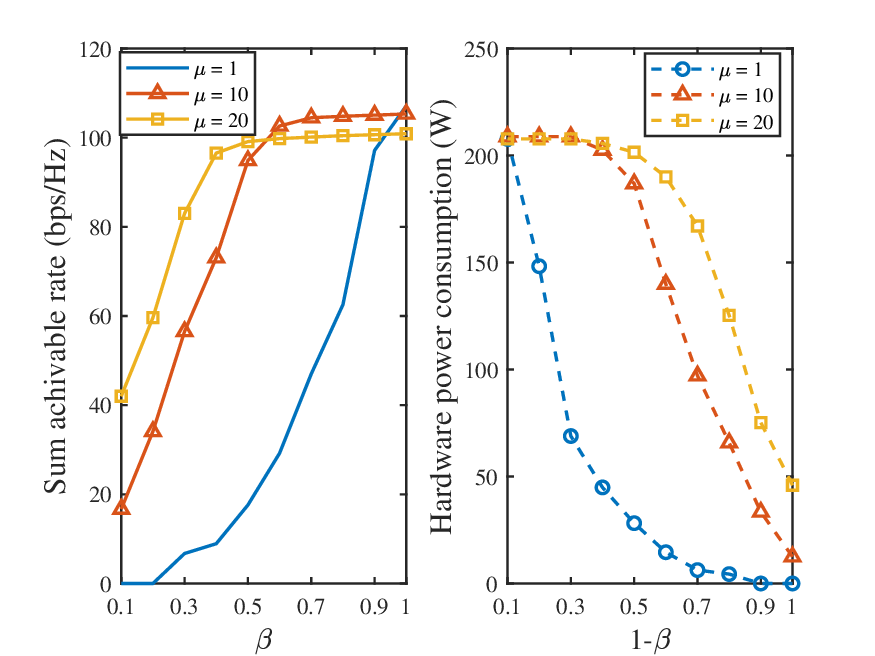}
 \label{Fig.8b}
}\caption{Performance tradeoff between the communication rate and hardware power consumption.}
\label{Fig.8}
\end{figure}

From Fig. \ref{Fig.8a}, It can be observed that increasing transmit power facilitates enlarging the rate-HPC tradeoff region. This can be explained by the fact that an increase in $P_{\text{max}}$ enables more information to be carried per data stream, thus leading to an increase in the communication rate. The rate increase in turn encourages the BS to activate more RF chains for data transmission, thus leading to a higher HPC. As shown in Fig. \ref{Fig.8b}, the achievable rate increases with the rise of the weight coefficient $\beta$ while the HPC degrades with the increasing coefficient $1-\beta$. It can be explained that with a large $\beta$, the objective function is predominated by the achievable rate and becomes tolerant to the HPC, so the WMMSE-TS algorithm tends to send more data streams for utilizing the multiplexing gains of near-field channels. In contrast, when we reduce $\beta$, the coefficient of HPC will be large. Thus, the proposed algorithm becomes sensitive to the increase in HPC, which prefers to decrease the number of transmitted data streams to avoid suffering high HPC. It can also be observed from Fig. \ref{Fig.8b} that decreasing the scaling coefficient $\mu$ is conducive to expanding the rate/HPC variation region. This is because $\mu$ directly impacts the data stream selection, i.e., equation \eqref{31}, where a small $\mu$ renders the WMMSE-TS algorithm more sensitive to the HPC, therefore resulting in a large rate-HPC tradeoff region.

\begin{figure}[t]
  \centering
  \includegraphics[scale = 0.52]{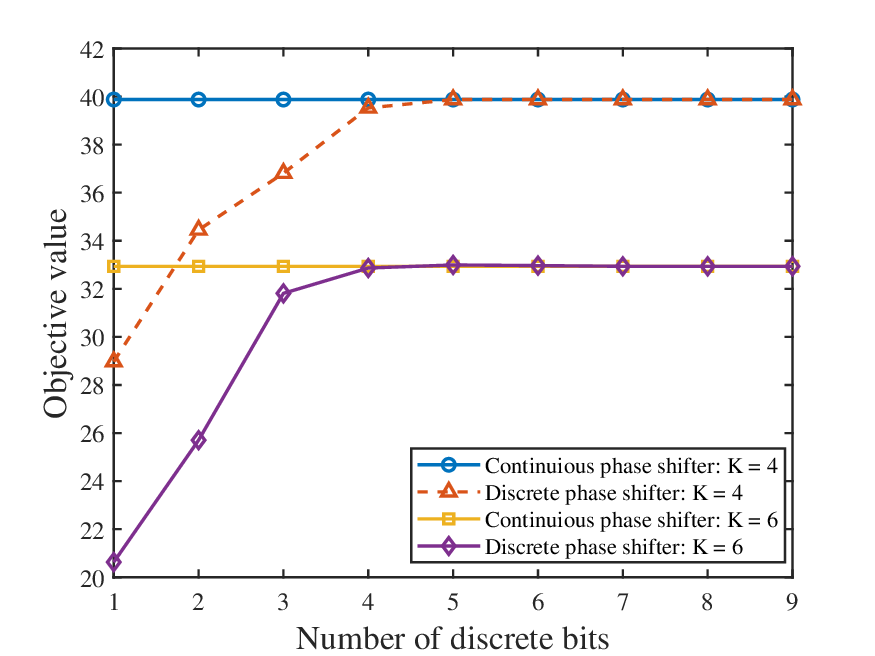}
  \caption{The performance of the PLI algorithm versus the number of discrete bits  with $P_{\text{max}}$ = 15 dBm.}
  \label{Fig.9}
\end{figure}
In Fig. \ref{Fig.9}, the objective value obtained by the PLI algorithm versus the number of the discrete phase-shift bits is illustrated. Firstly, we can observe from Fig. \ref{Fig.9} that objective value shows an upward trend with the rise of a number of discrete phase-shift bits, which is because increasing phase-shift bits benefits improving phase-shift resolution and reduces the quantification error in the PLI algorithm. Also, it can be found that a 4-bit discrete phase shifter can achieve comparable performance to the ideally continuous phase shifter, which is different from the conventional hybrid beamforming architecture that 3-bit discrete phase shifters are enough to achieve the near-optimal fully-digital beamforming \cite{PDD}. This phenomenon can be explained as follows. In the conventional hybrid beamforming structure, the feasible region of the analog beamformer lies on the boundary of a circle with radius 1 in the complex plane. However, in the dynamic hybrid beamforming, the feasible region of the analog beamformer extends to the interior of a circle of radius 2 in the complex plane, which thus requires more discrete phase-shift bits to approximate the continuous phase shifters. In addition, it can be observed that the objective value degrades when the BS supports more users. This is due to the fact that near-field communications suffer a more severe inter-user interference with an increased number of users, which deteriorates the communication performance of the network.

\section{Conclusion}\label{Section_6}
A dynamic hybrid beamforming architecture for near-field MIMO networks was proposed, where a switching module was introduced to strike a balance between the communication rates and HPC. An optimization problem of maximizing achievable communication rates while minimizing the HPC at the BS was formulated. For continuous phase shifters, a WMMSE-TS algorithm was proposed, which achieves the same performance as the fully-digital beamforming with the number of RF chains equaling that of the data streams. For discrete phase shifters, a PLI algorithm was proposed, where the analog beamformer and baseband digital beamformer are obtained in the closed-form expressions. Numerical results also unveiled that: 1) the proposed dynamic hybrid beamforming architecture outperforms the other baseline schemes; 2) the proposed algorithms can provide a flexible spatial multiplexing-power consumption tradeoff; 3) the 4-bit discrete phase shifters are sufficient to exhibit performance as good as continuous phase shifters.

\section*{Appendix A: Proof of Proposition \ref{Proposition_1}}
With the optimized fully-digital beamformer $\mathbf{\bar{W}}\in\mathbb{C}^{M_{\text{t}}\times KM_{\text{r}}}$, we carry out the QR decomposition to $\mathbf{\bar{W}}$, which is given by
\begin{align}\tag{A-1}\label{A-1}
\mathbf{\bar{W}} \overset{\text{QR}}{=} \mathbf{\tilde{V}}_{1}\mathbf{\tilde{W}}_{1},
\end{align}
where $\mathbf{\tilde{V}}_{1}\in\mathbb{C}^{M_{\text{t}}\times T_{\text{s}}}$ is a semi-unitary matrix and $\mathbf{\tilde{W}}_{1}\in\mathbb{C}^{T_{\text{s}}\times K M_{\text{r}}}$ is an upper triangular matrix. Note that $T_{k}$ denotes the rank of $\mathbf{\bar{W}}_{k}$, which satisfies $0\leq T_{\text{s}} \leq KM_{\text{r}}$. To proceed, let $\mathbf{\tilde{V}}_{2}$ denotes the conjugate transpose of $\mathbf{\tilde{V}}_{1}$, and perform the QR decomposition to $\mathbf{\tilde{V}}_{2}$, i.e., $\mathbf{\tilde{V}}_{2}\overset{\text{QR}}{=}\mathbf{\tilde{W}}_{2}\mathbf{\tilde{P}}$, \eqref{A-1} can be rewritten as
\begin{align}\tag{A-2}\label{A-2}
\mathbf{\bar{W}} \overset{\text{QR}}{=} \mathbf{\tilde{P}}^{H}\mathbf{\tilde{W}}_{2}^{H}\mathbf{\tilde{W}}_{1},
\end{align}
To guarantee the analog beamformer can be equivalently expressed by the sum of two unit-modulus phase shifters, $\mathbf{\tilde{P}}^{H}$ is further decomposed to $\mathbf{\tilde{P}}^{H}=\mathbf{P}\mathbf{\Xi}$, where $\mathbf{\Xi}\in\mathbb{C}^{T_{\text{s}}\times T_{\text{s}}}$ is a diagonal matrix with satisfying
\begin{align}\tag{A-3}\label{A-3}
\mathbf{\Xi}^{[i,i]} = \frac{\lceil\mathbf{p}_{i}\rceil}{2},
\end{align}
where $\mathbf{P}=[\mathbf{p}_{1},\cdots,\mathbf{p}_{T_{\text{s}}}]\in\mathbb{C}^{M_{\text{t}}\times T_{\text{s}}}$ and $\lceil\mathbf{p}_{i}\rceil$ denotes the modulus of elements in $\mathbf{p}_{i}$ that possesses the maximum modulus. Then, the optimal hybrid beamformer can be derived by \eqref{17} in Proposition \ref{Proposition_1}, where the elements of the analog beamformer have the maximum amplitude of $2$. According to Euler's formula, we have the identities below
\begin{align} \nonumber
z = &\left[\cos\left(\cos^{-1}\left(\frac{z}{2}\right)\right)+\cos\left(-\cos^{-1}\left(\frac{z}{2}\right)\right)\right]+\\ \nonumber
&\left[\jmath\sin\left(\cos^{-1}\left(\frac{z}{2}\right)\right)+\jmath\sin\left(-\cos^{-1}\left(\frac{z}{2}\right)\right)\right],\\ \tag{A-4}\label{A-4}
= &e^{\jmath\cos^{-1}\left(\frac{z}{2}\right)} + e^{-\jmath\cos^{-1}\left(\frac{z}{2}\right)},
\end{align}
\begin{align} \nonumber
\jmath z = &\left[\cos\left(\sin^{-1}\left(\frac{z}{2}\right)\right)+\cos\left(\pi-\sin^{-1}\left(\frac{z}{2}\right)\right)\right]+\\ \nonumber
&\left[\jmath\sin\left(\sin^{-1}\left(\frac{z}{2}\right)\right)+\jmath\sin\left(\pi-\sin^{-1}\left(\frac{z}{2}\right)\right)\right],\\ \tag{A-5}\label{A-5}
= &e^{\jmath\sin^{-1}\left(\frac{z}{2}\right)} + e^{-\jmath\left(\pi-\sin^{-1}\left(\frac{z}{2}\right)\right)},
\end{align}
where the real number $z$ satisfies $-2\leq z\leq 2$. Thus, the analog beamformer of \eqref{18} in Proposition \ref{Proposition_1} can be obtained. This completes the proof.

\end{document}